\documentclass{aa}  
\pdfoutput=0

\usepackage{graphicx}
\usepackage{txfonts}
\usepackage{upgreek}
\usepackage{color}
\usepackage{multirow}
\usepackage{subfig}
\usepackage{lscape}
\usepackage{ifthen}

\usepackage{natbib}
\bibpunct{(}{)}{;}{a}{}{,} 

\makeatletter
\DeclareRobustCommand*\textsubscript[1]{%
    \@textsubscript{\selectfont#1}}
  \newcommand{\@textsubscript}[1]{%
    {\m@th\ensuremath{_{\mbox{\fontsize\sf@size\z@#1}}}}}
\makeatother

\begin{document}
   \title{Discovering young stars in the Gum\,31 region with infrared observations\thanks{This work is based in part on data collected by \textit{Herschel}, an ESA space observatory with science instruments provided by European-led Principal Investigator consortia and with important participation from NASA, and on data observed by VISTA (ESO project number 088.C-0117(A)), an ESO survey telescope developed by a consortium of 18 universities in the United Kingdom, led by Queen Mary, University of London.}}

   \author{H. Ohlendorf
          \inst{1}
           \and
           T. Preibisch
           \inst{1}
           \and
           B. Gaczkowski
           \inst{1}
           \and
           T. Ratzka
           \inst{1}
           \and
           J. Ngoumou
           \inst{1}
           \and
           V. Roccatagliata
           \inst{1}
           \and
           R. Grellmann
           \inst{1}
          }

   \institute{Universit\"ats-Sternwarte M\"unchen, Ludwig-Maximilians-Universit\"at,
             Scheinerstr. 1, 81679 M\"unchen, Germany \\
              \email{ohlendorf@usm.uni-muenchen.de}}

   \date{Received 14 August, 2012; accepted 24 January, 2013}

\abstract
   {The Gum\,31 bubble containing the stellar cluster NGC\,3324 is a poorly-studied young region close to the Carina Nebula.}
   {We are aiming to characterise the young stellar and protostellar population in and around Gum\,31 and to investigate the star-formation process in this region.}
   {We identify candidate young stellar objects from \textit{Spitzer}, WISE, and \textit{Herschel} data. Combining these, we analyse the spectral energy distributions of the candidate young stellar objects. With density and temperature maps obtained from \textit{Herschel} data and comparisons to a `collect and collapse' scenario for the region we are able to further constrain the characteristics of the region as a whole.}
   {661 candidate young stellar objects are found from WISE data, 91 protostar candidates are detected through \textit{Herschel} observations in a $1.0\degr \times 1.1\degr$ area. Most of these objects are found in small clusters or are well aligned with the H\,II bubble. We also identify the sources of Herbig-Haro jets. The infrared morphology of the region suggests that it is part of the larger Carina Nebula complex.}
   {The location of the candidate young stellar objects in the rim of the H\,II bubble is suggestive of their being triggered by a `collect and collapse' scenario, which agrees well with the observed parameters of the region. Some candidate young stellar objects are found in the heads of pillars, which points towards radiative triggering of star formation. Thus, we find evidence that in the region different mechanisms of triggered star formation are at work. Correcting the number of candidate young stellar objects for contamination we find $\sim 600$ young stellar objects in Gum\,31 above our completeness limit of about 1\,\ensuremath{M_\odot}. Extrapolating the intital mass function down to 0.1\,\ensuremath{M_\odot}, we estimate a total population of $\sim 5000$ young stars for the region.}


  \maketitle
%

\section{Introduction}
\label{sec:intro}

The bubble-shaped H\,II region Gum\,31 around the young stellar cluster NGC\,3324 is located $\approx 1\degr$ north-west of the Carina Nebula (NGC\,3372; see \citeauthor{smith2008} \citeyear{smith2008} for a recent review). While numerous observations of the Carina Nebula have been performed in the last few years and provided comprehensive information about the stellar populations as well as the cloud properties \citep{yonekura2005, smith2007_census, kramer2008, smith2010_jets, smith2010_spitzer, townsley2011, preibisch2011_cccp, preibisch2011_hawki, preibisch2011_laboca, preibisch2012, salatino2012} the Gum\,31 region has not received much attention.
Despite its interesting morphology and the publicity of HST images of its western rim (Hubble News Release STScI-2008-34), the H\,II region and its stellar population remain rather poorly studied until today. This seems to be related to its celestial position: the closeness to the extremely eye-catching Carina Nebula has always overshadowed NGC\,3324.

The physical relation between NGC\,3324 and the Carina Nebula complex (CNC) is still unclear. Recent distance determinations of Gum\,31 and NGC\,3324 yielded values of 2.3\,\ensuremath{\mathrm{kpc}} for the cluster NGC\,3324 \citep[Catalogue of Open Cluster Data;][]{kharchenko2005} and $2.5\pm 0.3$\,\ensuremath{\mathrm{kpc}} for the clouds in and around Gum\,31 \citep{barnes2010}. This implies that the NGC\,3324/Gum\,31 region is located at the same distance as the Carina Nebula \citep[$\approx 2.3$\,\ensuremath{\mathrm{kpc}}; see][]{smith2006_homunculus}.

The diameter of the Gum\,31 H\,II region is $\sim 15\arcmin$ (10\,\ensuremath{\mathrm{pc}}). This H\,II region in turn is surrounded by an expanding H\,I shell which encloses it almost completely \citep{cappa2008}.

The available information about the stellar population of NGC\,3324 is restricted to the three brightest stars.
The brightest star of the cluster is the multiple star HD\,92206, which is constituted by two O6.5V~stars HD\,92206\,A and B and O8.5V~star HD\,92206\,C \citep{maizApellaniz2004}.
A further very luminous star in the region is the A0 supergiant HD\,92207 (= V370 Car, $V = 5.49$). This object is a very luminous ($\log \left(L/\,\ensuremath{L_\odot} \right) = 5.56$) massive ($M_{\rm initial} \approx 30\,\ensuremath{M_\odot}$) evolved star \citep[age $\approx 7\pm 1$\,\ensuremath{\mathrm{Myr}}; ][]{przybilla2006} that drives a very strong stellar wind \citep[$\dot{M} = 1.3 \cdot 10^{-6}\,\ensuremath{M_\odot}\,{\rm yr}^{-1}$;][]{kudritzki1999}. The membership of HD\,92207 is not entirely clear: \citet{claria1977} and \citet{carraro2001} assume it not to be part of the cluster. However, \citet[][from HD\,92207 being wrapped in a nebular shell associated with Gum\,31]{forte1976} and \citet[][from proper motions]{baumgardt2000} determine it to be a member.

\citet{carraro2001} identify 25 candidate members of NGC\,3324 with optical photometry and suggest that the cluster is very young ($\lesssim 2$ -- 3\,\ensuremath{\mathrm{Myr}}). With three O-type stars ($M \ge 18\,\ensuremath{M_\odot}$), the field star initial mass function (IMF) representation by \citet{kroupa2002} suggests that there should be $\approx 1500$ low-mass ($0.1\,\ensuremath{M_\odot} \le M \le 2\,\ensuremath{M_\odot}$) stars present. This implies that the vast majority of the stellar population of NGC\,3324 is still unknown.

In addition to the optically visible stellar cluster in the H\,II region, there is a population of young stars embedded in the molecular cloud surrounding the H\,II region. The stars in this population are only seen in infrared images \citep{cappa2008} or traced by their protostellar jets \citep{smith2010_jets}. \citet{cappa2008} sampled point sources in a region of 20\arcmin\ radius centred on NGC\,3324 from the IRAS, MSX and 2MASS point-source catalogues and identify 12 (IRAS), 9 (MSX) and 26 (2MASS) candidate young stellar objects (cYSOs) using colour-colour criteria. Due to the limited sensitivity and angular resolution of these data, the currently known few dozen of embedded infrared sources represent only the tip of the iceberg; many more embedded young stellar objects (YSOs) must be present in this area and waiting to be discovered.

In this paper, we are aiming to characterise the protostellar and young stellar population of the stellar cluster, the surrounding H\,II region and its environs. Our study is based on \textit{Spitzer}, WISE, \textit{Herschel}, and VISTA data, which provide much better sensitivity and angular resolution than the previously existing data sets.

We will describe the data sets we used in Sect.~\ref{sec:data}. In Sect.~\ref{sec:morphology}, we will describe the general morphology of the infrared clouds in and around the Gum\,31 region and its implications for our work. Detections of cYSOs from the IR data, including the identification method, and their spatial distribution will be described in Sect.~\ref{sec:ysos}. Furthermore, we have analysed point-like sources as detected with both \textit{Herschel} and \textit{Spitzer} and derived their spectral energy distributions (SEDs; Sect.~\ref{sec:herschelspitzer}). We have also identified likely sources to two previously detected Herbig-Haro jets (Sect.~\ref{sec:hhjets}). Inferences will be discussed in Sect.~\ref{sec:conclusions}.

\begin{figure*}
\centering
\sidecaption
    \includegraphics[width=12cm]{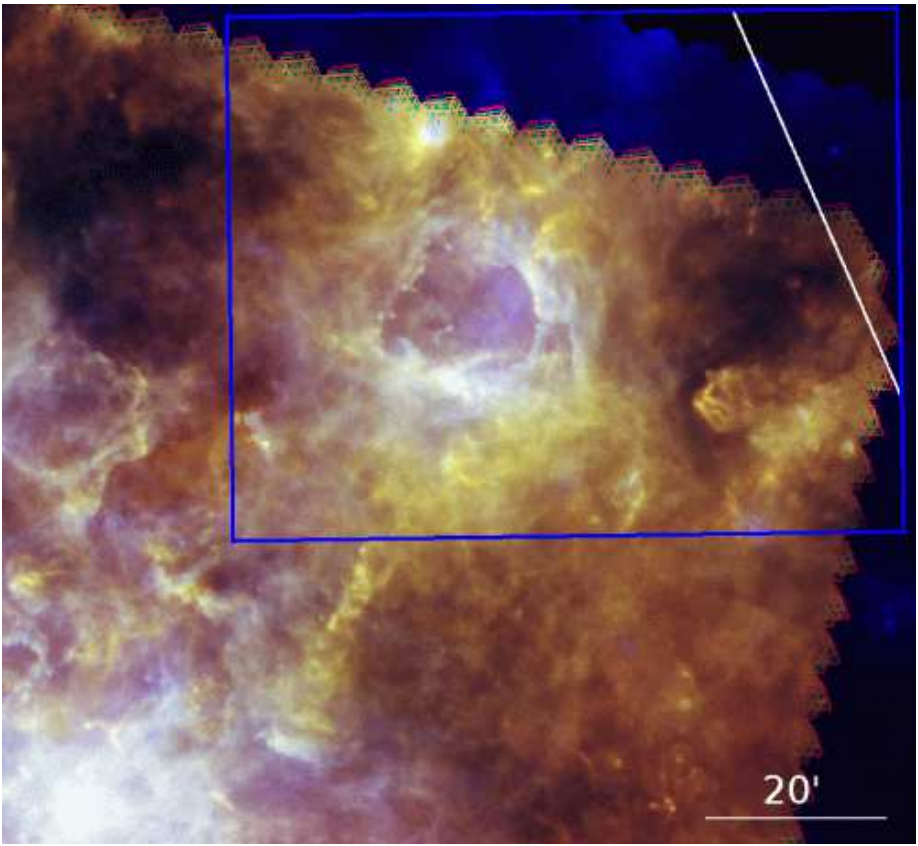}
   \caption{The Gum\,31 region and its connection to the central Carina Nebula, seen in a \textit{Herschel} RGB image, with the PACS 70\,\ensuremath{\upmu\mathrm{m}} image in blue, SPIRE 250\,\ensuremath{\upmu\mathrm{m}} in green, and SPIRE 500\,\ensuremath{\upmu\mathrm{m}} in red. The blue box marks the $1.1\degr \times 1.0\degr$ area used for analysis here. The diagonal white line marks the border to the left of which we obtained IRAC photometric data.}
  \label{fig:spitzer_herschel_ausschnitt}
\end{figure*}

\begin{figure*}
\centering
\includegraphics[width=\textwidth]{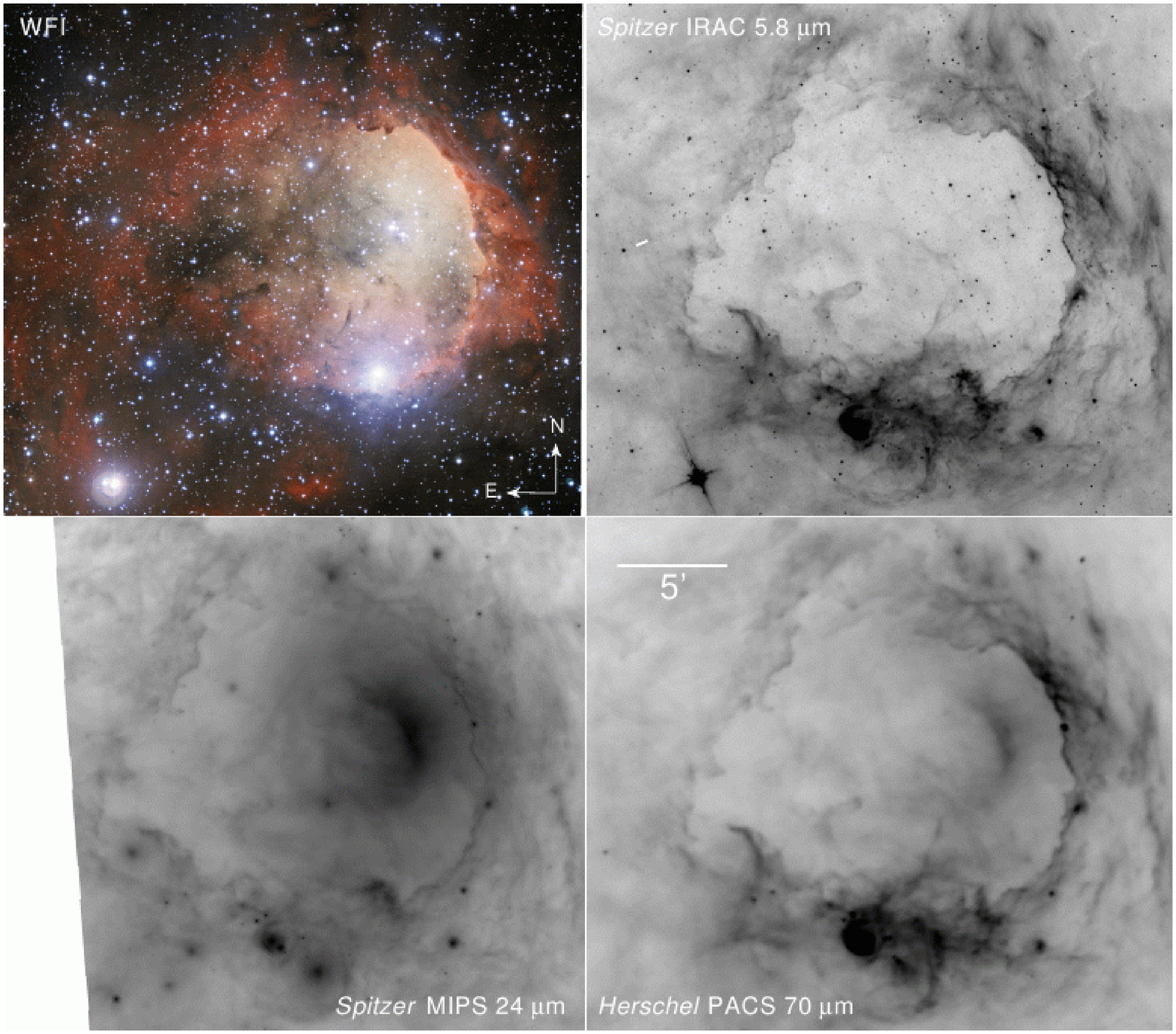}
  \caption{An overview of the Gum\,31 region in the optical, mid-IR, and far-IR. The ESO Wide Field Imager (WFI) optical image [ESO image release eso1207] shows the \emph{V}~band in blue, \emph{R}~band in yellow, O\,III 501\,nm emission in green, and H$\upalpha$ in red.}
  \label{fig:wfi_irac_mips}
\end{figure*}

\section{Observational data}
\label{sec:data}

\subsection{\textit{Spitzer} images and photometry}
\label{sec:data-spitzer}

\subsubsection{IRAC}
\label{sec:data-spitzer-irac}

We are using data taken during the \textit{Spitzer} cold mission phase, in July 2008, with the InfraRed Array Camera \citep[IRAC;][]{fazio2004} (``Galactic Structure and Star Formation in Vela-Carina'' programme; PI: Steven R. Majewski, Prog-ID: 40791); retrieved through the \textit{Spitzer} Heritage Archive.\footnote{An overview of the observations used can be found in Table~\ref{tab:obslog}.} The Basic Calibrated Data were assembled into mosaics using version 18.4.9 of the MOsaicker and Point source EXtractor package \citep[MOPEX;][]{makovoz2005} provided by the \textit{Spitzer} Science Centre\footnote{\url{http://irsa.ipac.caltech.edu/}} (SSC). From the complete data set recorded in the observation programme, we use only a $1.0\degr \times 1.1\degr$ area corresponding to the Gum\,31 region and a small region to the west of it. This complements the area as described by \citet{gaczkowski2012} to complete the area surveyed with our \textit{Herschel} observation programme (cf.\ Sect.~\ref{sec:data-herschel}).

\begin{table*}
\caption[]{Overview of observations with \textit{Spitzer} and \textit{Herschel} used in this work.}
\label{tab:obslog}
\centering
\begin{tabular}{lllrll}
\hline\hline
\noalign{\smallskip}
Instrument & AOR & Observation mode & Observation time & Date & PI  \\
\noalign{\smallskip}
\hline
\noalign{\smallskip}

\textit{Spitzer} IRAC   & 23708160        & IRAC map        & 3301\,s    & 19 July 2008     & S. R. Majewski \\
\textit{Spitzer} IRAC   & 23699200        & IRAC map        & 3057\,s    & 19 July 2008     & S. R. Majewski \\
\textit{Spitzer} IRAC   & 23704320        & IRAC map        & 3070\,s    & 19 July 2008     & S. R. Majewski \\
\textit{Spitzer} IRAC   & 23688192        & IRAC map        & 3082\,s    & 19 July 2008     & S. R. Majewski \\
\textit{Spitzer} IRAC   & 23701504        & IRAC map        & 3086\,s    & 19 July 2008     & S. R. Majewski \\
\textit{Spitzer} IRAC   & 23695360        & IRAC map        & 3086\,s    & 19 July 2008     & S. R. Majewski \\
\textit{Spitzer} IRAC   & 23696896        & IRAC map        & 3087\,s    & 19 July 2008     & S. R. Majewski \\
\textit{Spitzer} IRAC   & 23706368        & IRAC map        & 3088\,s    & 19 July 2008     & S. R. Majewski \\
\textit{Spitzer} IRAC   & 23684352        & IRAC map        & 3312\,s    & 19 July 2008     & S. R. Majewski \\

\textit{Spitzer} MIPS   & 15054080        & MIPS phot       & 9469\,s    & 12 June 2006     & J. Hester      \\

\textit{Herschel} PACS  & 1342211615      & SpirePacsParallel & 11889\,s   & 26 December 2010 & T. Preibisch   \\
\textit{Herschel} PACS  & 1342211616      & SpirePacsParallel & 12863\,s   & 26 December 2010 & T. Preibisch   \\
\textit{Herschel} SPIRE & 1342211615      & SpirePacsParallel & 11889\,s   & 26 December 2010 & T. Preibisch   \\
\textit{Herschel} SPIRE & 1342211616      & SpirePacsParallel & 12863\,s   & 26 December 2010 & T. Preibisch   \\

\hline
\end{tabular}
\end{table*}

Using the Astronomical Point source EXtractor (APEX) module of MOPEX we then performed source detection and photometry on these IRAC mosaics. Photometry was first carried out individually for each image in the stack and subsequently combined internally to provide photometry data for the entire mosaic. Sets of Point Response Functions (PRFs) available from the SSC, chosen appropriately for the time of the observations, were employed.
Before the mosaics were constructed, outliers were removed using the Box Outlier Detection method within MOPEX and backgrounds between the tiles were matched using the Overlap module. Subsequently, three \textit{Spitzer} Astronomical Observation Requests (AORs) each were merged into a single mosaic.

PRF-fitting photometry was carried out separately for all four IRAC bands (3.6\,\ensuremath{\upmu\mathrm{m}}, 4.5\,\ensuremath{\upmu\mathrm{m}}, 5.8\,\ensuremath{\upmu\mathrm{m}} and 8.0\,\ensuremath{\upmu\mathrm{m}}). 
This photometric information was then combined into a single catalogue by a simple nearest-neighbour matching algorithm, taking the 4.5\,\ensuremath{\upmu\mathrm{m}} band as reference. (We chose this band as our reference band because it clearly is the most sensitive one and less likely to be subject to misidentifications. These are common especially in the longest-wavelength bands where random fluctuations in nebulosities are detected as point-like sources.) Thus, the resulting catalogue is rather conservative and excludes any object not detected in the 4.5\,\ensuremath{\upmu\mathrm{m}} band. On the other hand, it minimises contamination. The total number of point-like sources detected in at least one band in the study area is 57\,828.

We noticed a slight misalignment between the \textit{Spitzer} bands and so globally shifted the 4.5\,\ensuremath{\upmu\mathrm{m}}, 5.8\,\ensuremath{\upmu\mathrm{m}}, and 8.0\,\ensuremath{\upmu\mathrm{m}} band positions by $-0.17\arcsec$, $-0.1\arcsec$, and $-0.13\arcsec$, respectively, in declination with regard to the 3.6\,\ensuremath{\upmu\mathrm{m}} band that we found to be well aligned with the Two Micron All Sky Survey \citep[2MASS;][]{skrutskie2006}. In the resulting catalogue, the root mean square (RMS) of the spatial deviations between the IRAC and the 2MASS position is 0.22\arcsec.

Due to the very strong spatial inhomogeneity of the cloud emission in our maps, the sensitivity for IRAC cannot be precisely quantified by a single value. Instead, we characterise it by two typical values, the detection limit and the typical completeness limit. The former is quantified by the faintest sources in our sample. For 3.6\,\ensuremath{\upmu\mathrm{m}}, 4.5\,\ensuremath{\upmu\mathrm{m}}, 5.8\,\ensuremath{\upmu\mathrm{m}}, and 8.0\,\ensuremath{\upmu\mathrm{m}} these have fluxes of 100$\,\upmu$Jy, 100$\,\upmu$Jy, 330$\,\upmu$Jy, and 280$\,\upmu$Jy, respectively.

\begin{figure*}
\centering
   \begin{minipage}{0.49\textwidth}
    \includegraphics[width=\textwidth]{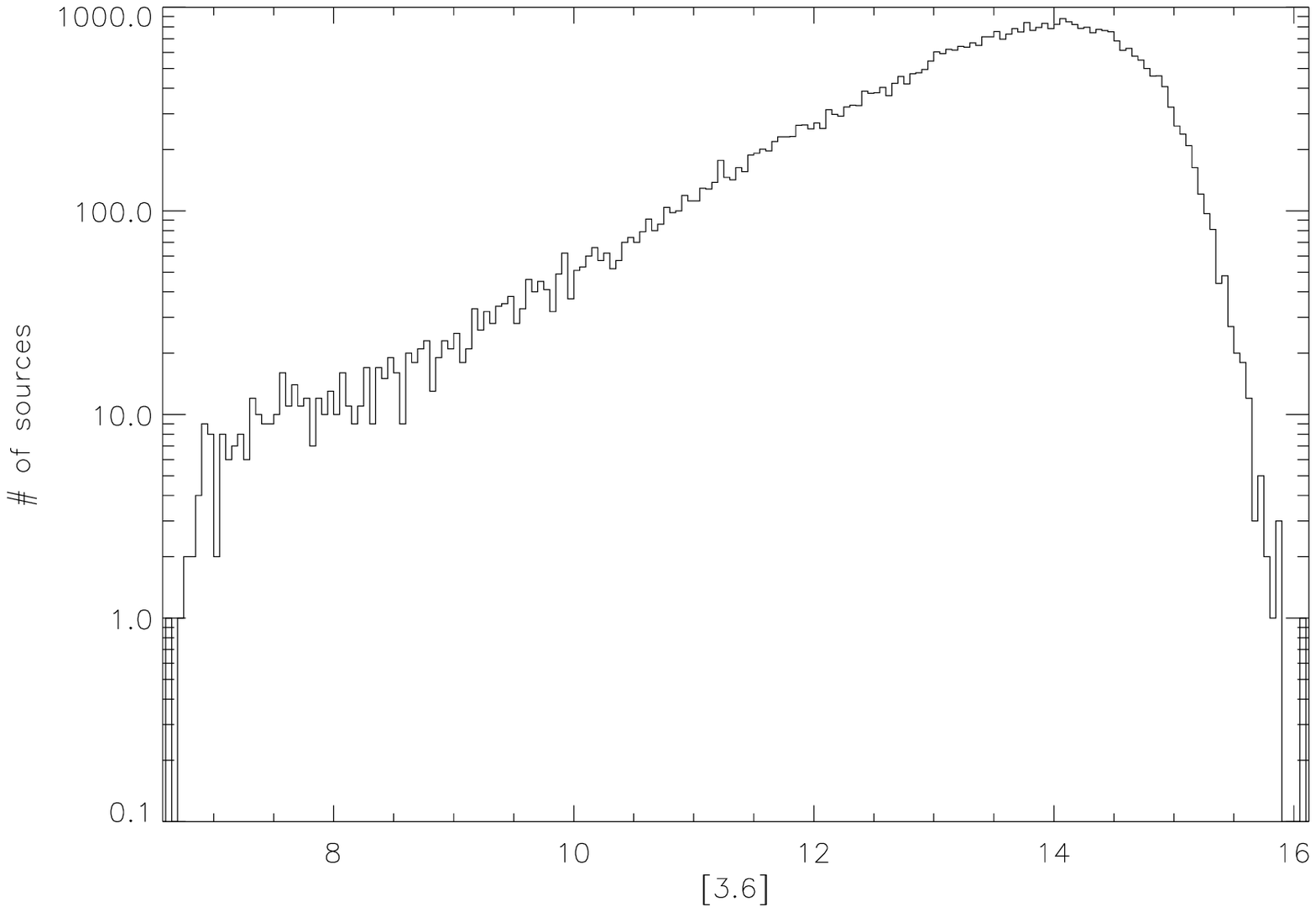}
   \end{minipage}
  \begin{minipage}{0.49\textwidth}
    \includegraphics[width=\textwidth]{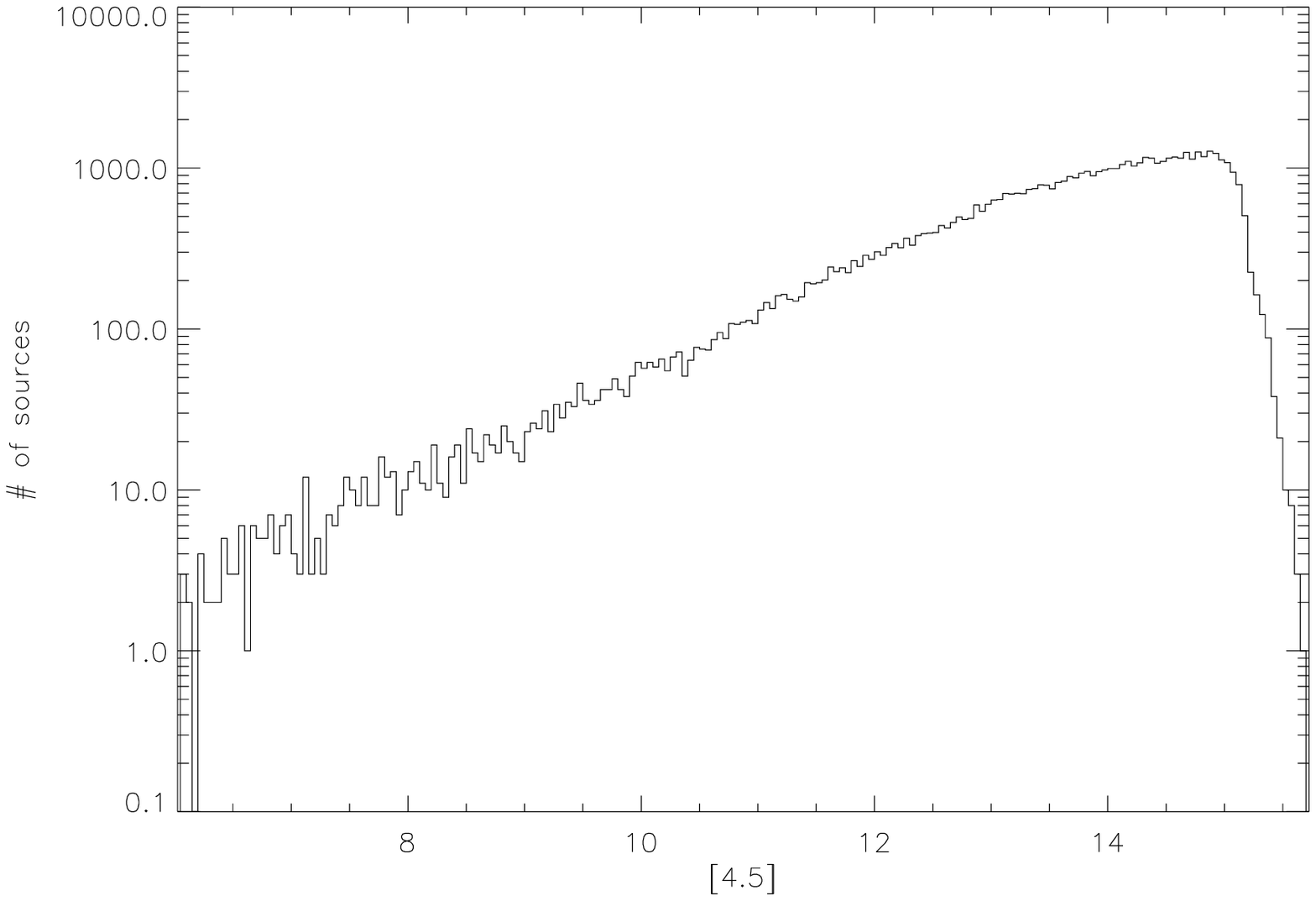}
   \end{minipage}
  \begin{minipage}{0.49\textwidth}
    \includegraphics[width=\textwidth]{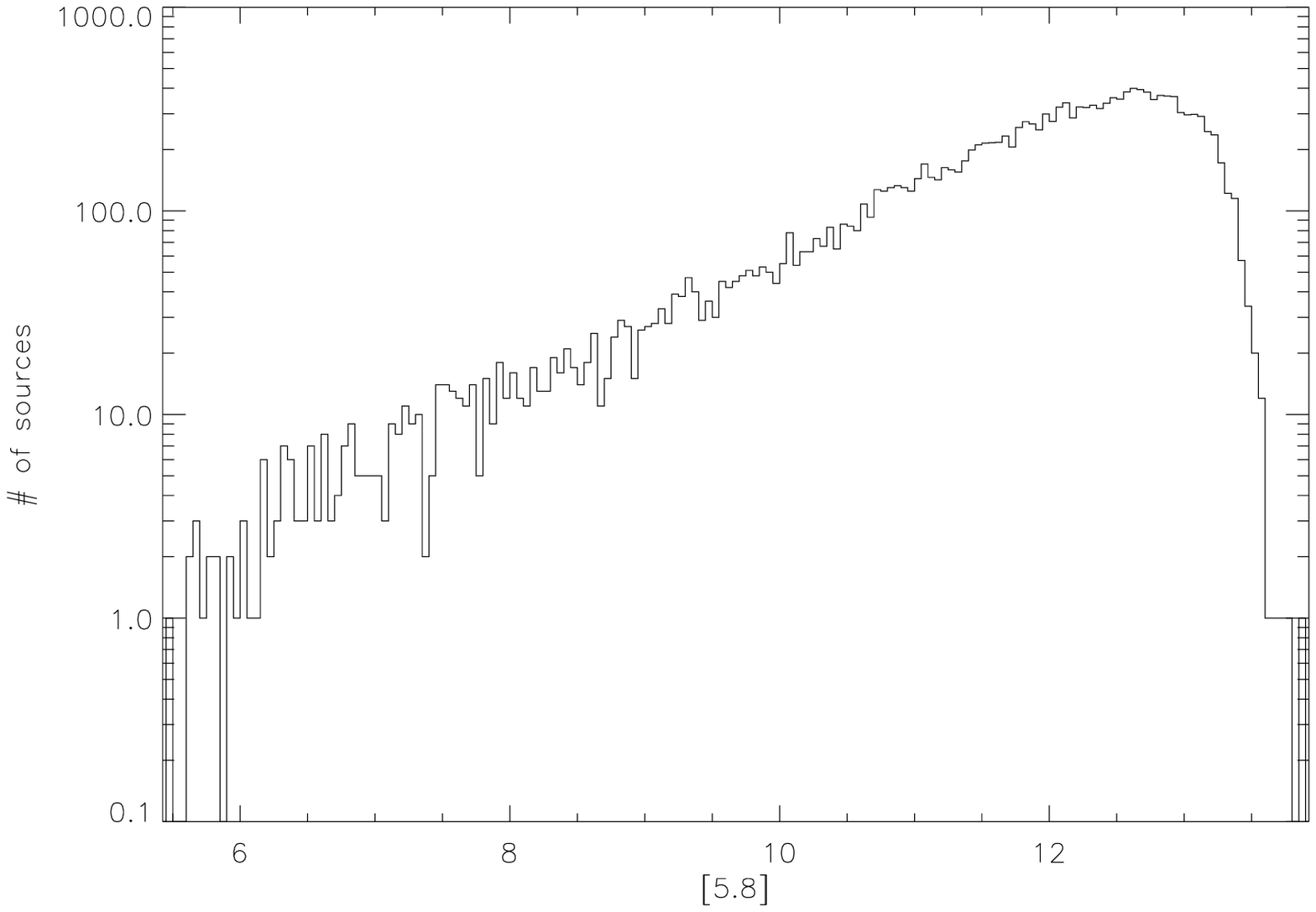}
   \end{minipage}
  \begin{minipage}{0.49\textwidth}
    \includegraphics[width=\textwidth]{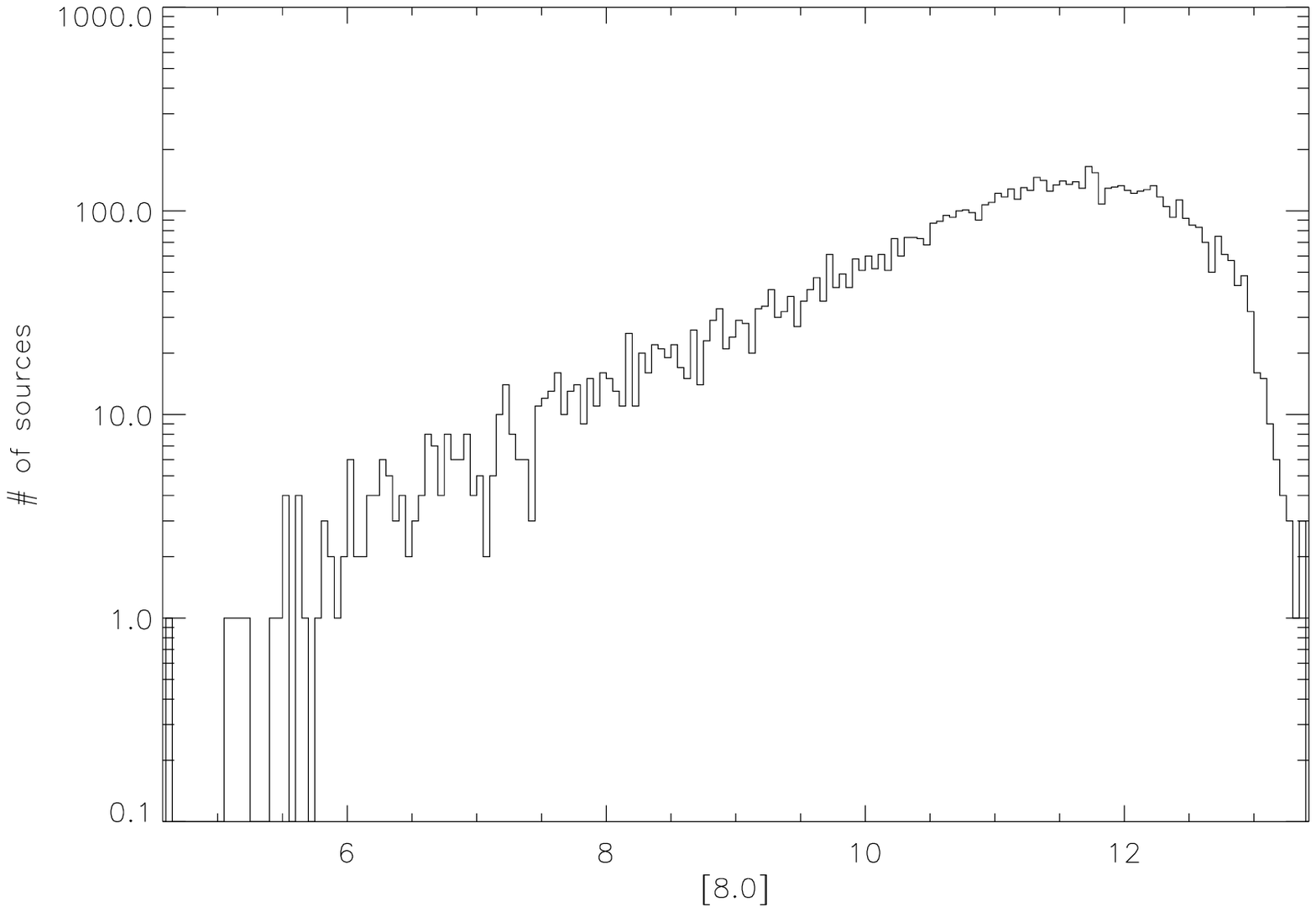}
   \end{minipage}
  \caption{Histograms of the measured source magnitudes in our \textit{Spitzer} IRAC catalogue.}
  \label{fig:spitzer_histograms}
\end{figure*}

To estimate the completeness limits, we constructed histograms of the measured source magnitudes (Fig.~\ref{fig:spitzer_histograms}) and estimate the point where the rise in the source count is no longer well-described by a power-law. Although it is not a formal measure of completeness, the turnover in source count curves can serve as a proxy to show the typical values of the completeness limit across the field. In this way we estimate completeness limits of $\approx 1.5$\,mJy, $\approx 0.7$\,mJy, $\approx 1.2$\,mJy and $\approx 1.6$\,mJy for 3.6\,\ensuremath{\upmu\mathrm{m}}, 4.5\,\ensuremath{\upmu\mathrm{m}}, 5.8\,\ensuremath{\upmu\mathrm{m}}, and 8.0\,\ensuremath{\upmu\mathrm{m}}, respectively.

Comparing these to numerical models of stellar evolution \citep{baraffe1998}, we find that for an age of $\sim 3$\,Myr the photospheric emission of a 1\,\ensuremath{M_\odot} star is well above the detection limit for WISE (Sect.~\ref{sec:data-wise}) in the 3.4\,\ensuremath{\upmu\mathrm{m}} and 4.6\,\ensuremath{\upmu\mathrm{m}} bands while for IRAC the sensitivity extends down to $\sim 0.5$\,\ensuremath{M_\odot}. For younger stars this boundary shifts upwards so that with WISE 1\,\ensuremath{\mathrm{Myr}} old 0.5\,\ensuremath{M_\odot} stars are still detectable while with IRAC we could reach down to $\sim 0.25$\,\ensuremath{M_\odot}. 
If we compare the completeness limits to the RADMC model \citep{dullemond2004} which provides calculations of continuum radiative transfer in axisymmetric circumstellar dust distributions around a central illuminating star, we find that disk masses as low as 0.013\,\ensuremath{M_\odot} would still be detectable for a 1\,\ensuremath{M_\odot} YSO.

\subsubsection{MIPS}
\label{sec:data-spitzer-mips}

For the detailed analysis in Sect.~\ref{sec:herschelspitzer} the IRAC data were complemented by photometry on a Multiband Imaging Photometer for \textit{Spitzer} (MIPS) \citep{rieke2004} 24\,\ensuremath{\upmu\mathrm{m}} map (``\textit{Spitzer} Follow-up of HST Observations of Star Formation in H\,II Regions'' programme; PI: Jeff Hester, Prog-ID: 20726) retrieved through the \textit{Spitzer} Heritage Archive. The MIPS images were searched by eye for point-like sources coinciding with the previously identified IRAC and \textit{Herschel} (Sect.~\ref{sec:data-herschel}) sources. As this field is only sparsely populated with point-like sources in the 24\,\ensuremath{\upmu\mathrm{m}} band, they were easily found by eye. MIPS has a PSF of FWHM 6\arcsec in the 24\,\ensuremath{\upmu\mathrm{m}} band, slightly degrading towards the edges of the $5.4\arcmin \times 5.4\arcmin$ field of view, sampled onto pixels with a size of 2.55\arcsec. According to the \href{http://irsa.ipac.caltech.edu/data/SPITZER/docs/mips/mipsinstrumenthandbook}{MIPS Instrument Handbook}, the 24\,\ensuremath{\upmu\mathrm{m}} sensitivity is highly dependent on the observed sky region, in an ideal case it is expected to be $\approx 35\upmu$Jy.

Source detection and point-spread function (PSF) photometry were then performed with StarFinder \citep{diolaiti2000}.\footnote{\url{http://www.bo.astro.it/StarFinder/index.htm}} The PSF is derived as a median value from four point sources. It is then fit to each identified point source together with a slanting plane for the background and taking into account the contribution of bright stars adjacent to the fitting region. In this way the sub-sample of MIPS point-like sources defined by IRAC and \textit{Herschel} detection is analysed from the brightest (4570\,mJy) to the faintest (5\,mJy) objects.

\subsection{\textit{Herschel} far-infrared maps}
\label{sec:data-herschel}

We analysed the Gum\,31 region using data from our recent \textit{Herschel} far-infrared survey of the CNC.
These observations were performed in December 2010 (Open time project, PI: Thomas Preibisch, Prog-ID: OT1-tpreibis-1), using the parallel fast scan mode at $60 \arcsec \mathrm{s}^{-1}$. With simultaneous five-band imaging with PACS \citep{poglitsch2010} at 70\,\ensuremath{\upmu\mathrm{m}} and 160\,\ensuremath{\upmu\mathrm{m}} and SPIRE \citep{griffin2010} at 250\,\ensuremath{\upmu\mathrm{m}}, 350\,\ensuremath{\upmu\mathrm{m}}, and 500\,\ensuremath{\upmu\mathrm{m}} two orthogonal scan maps were obtained to cover an area of $2.3\degr \times 2.3 \degr$. The angular resolutions of the maps are 5\arcsec, 12\arcsec, 18\arcsec, 25\arcsec, and 36\arcsec\ for the 70\,\ensuremath{\upmu\mathrm{m}}, 160\,\ensuremath{\upmu\mathrm{m}}, 250\,\ensuremath{\upmu\mathrm{m}}, 350\,\ensuremath{\upmu\mathrm{m}}, and 500\,\ensuremath{\upmu\mathrm{m}} band, respectively. At a distance of 2.3\,\ensuremath{\mathrm{kpc}} this corresponds to physical scales from 0.06 to 0.4\,\ensuremath{\mathrm{pc}}.

A full description of these observations and the subsequent data processing is given by \citet{preibisch2012} and \citet{gaczkowski2012}. 
Detection and photometry of  point-like sources in \textit{Herschel} bands were carried out with CUTEX \citep[][also used for the Hi-GAL survey]{molinari2011}, a software package developed especially for maps with complex background. CUTEX calculates the second-order derivatives of the signal map in four directions (x, y and their diagonals), allowing the identification of point-like sources by their steep brightness gradients. A more detailed description of the photometry process is given by \citet{gaczkowski2012}.

As any source catalogue based on maps with strong and highly spatially inhomogeneous background emission, our \textit{Herschel} source lists will miss some faint
sources and at the same time contain a small number of spurious detections. To improve the reliability of the source catalogue, we restricted our analysis to those sources that are detected (independently) in at least two \textit{Herschel} maps and for the SED construction in Sect.~\ref{sec:herschelspitzer} even to those detected in at least three bands. For this, the point-like sources detected in each individual \textit{Herschel} band were matched. The matching process was described by \citet{gaczkowski2012}.
In the area, we detect 91 point-like sources in at least two bands and 59 in at least three bands. For each of them we checked coincidence with IRAC-identified point-like sources. We excluded all cases where either the spatial coincidence of \textit{Herschel} and IRAC source was not clear or where more than one IRAC source appeared as a possible counterpart. This results in the 16 sources identified both in \textit{Herschel} and in IRAC wavelengths that are subject to the analysis in Sect.~\ref{sec:herschelspitzer}.

The detection limit can be approximated by the fluxes of the faintest detected sources; this is in the range of $\approx 1$\,Jy to $\approx 2$\,Jy in our maps. An estimate for the `typical' completeness limit (i.\,e.\ the limit above which we can expect most objects in the survey area to be detected) can be derived from the histograms of the fluxes similar to those described in Sect.~\ref{sec:data-spitzer-irac}. The corresponding limits are at $\sim 10$\,Jy, $\sim 15$\,Jy, $\sim 10$\,Jy, $\sim 10$\,Jy, and $\sim 6$\,Jy for 70\,\ensuremath{\upmu\mathrm{m}}, 160\,\ensuremath{\upmu\mathrm{m}}, 250\,\ensuremath{\upmu\mathrm{m}}, 350\,\ensuremath{\upmu\mathrm{m}}, and 500\,\ensuremath{\upmu\mathrm{m}}, respectively.

\subsection{VISTA near-IR images}

An \emph{H}-band image of the area around Gum\,31 was obtained with the VISTA InfraRed CAMera (VIRCAM) \citep{dalton2006} in the night of 15 January 2012 as the first observation of our Visible and Infrared Survey Telescope for Astronomy (VISTA) \citep{irwin2004, emerson2010} survey of the Carina Nebula complex (ESO project number 088.C-0117(A)). VISTA is a 4-m class wide field survey telescope that provides a $1.3\degr \times 1.0\degr$ field of view. The near-infrared camera VIRCAM consists of an array of sixteen individual $2048 \times 2048$ pixel Raytheon VIRGO IR detectors, providing more than 67~million pixels with a nominal pixel size of 0.339\arcsec on the sky and sensitive in a wavelength range of 0.85 -- 2.4\,\ensuremath{\upmu\mathrm{m}}. 
Since the sixteen chips are non-contiguous they produce a set of sixteen non-contiguous images called a `pawprint'. For a contiguous sky coverage six pawprints, offset in x- and y-direction, are combined. The resulting `tile' covers an area of $1.5\degr \times 1.2\degr$ on the sky.

For each of the six pawprints, at five jitter positions 27 exposures with an integration time of 2\,s each have been obtained. As it turned out that the seeing conditions (with an average FWHM of $\approx 1.7\arcsec$) did not meet the pre-specified quality criteria, the observations were terminated
after the completion of this observing block. The observations in the other filters and for the rest of our Carina Nebula mosaic positions were successfully completed a few months later and processed by the VISTA Data Flow System at the Cambridge Astronomy Survey Unit. However, at the time of writing the photometric calibration of these data is still in progress. Therefore, we use only the image data for a (preliminary) scientific analysis here, but no photometric values.

A preliminary photometric calibration showed that objects as faint as $H \approx 18.5$ are clearly detectable in our VISTA image. This is about four magnitudes deeper than the nominal 2MASS Survey completeness limit for crowded locations near the galactic plane of $H \approx  14.5$ \citep{skrutskie2006}.

\subsection{WISE}
\label{sec:data-wise}

We used catalogue data from the Wide-field Infrared Survey Explorer \citep[WISE; ][]{wright2010} All-Sky Data Release \citep{cutri2012}. These data were taken during the WISE cold mission phase from January to August 2010. We used the standard aperture-corrected magnitudes obtained with apertures of 8.25\arcsec and corrections of 0.222\,mag, 0.280\,mag and 0.665\,mag for 3.4\,\ensuremath{\upmu\mathrm{m}}, 4.6\,\ensuremath{\upmu\mathrm{m}}, and 12\,\ensuremath{\upmu\mathrm{m}} \citep{cutriEtal2012_wiseExplanatorySupplement}.
The WISE observations have an angular resolution of 6.1\arcsec, 6.4\arcsec, 6.5\arcsec, and 12.0\arcsec\ for 3.4\,\ensuremath{\upmu\mathrm{m}}, 4.6\,\ensuremath{\upmu\mathrm{m}}, 12\,\ensuremath{\upmu\mathrm{m}}, and 22\,\ensuremath{\upmu\mathrm{m}}, respectively \citep{wright2010}. In our survey area the catalogue contains 20\,739 point sources with detection in at least one band. Source confusion in the catalogue should not constitute a problem in the analysis here as we required a matching radius of 0.6\arcsec\ to match WISE to IRAC detections, the same as we used for the IRAC inter-band matching. On this scale, we were able to distinguish clearly between nearest and second-nearest neighbours.

In way parallel to the way described in Sect.~\ref{sec:data-spitzer-irac}, we estimate the detection limits to be 90$\,\upmu$Jy, 70$\,\upmu$Jy and 700$\,\upmu$Jy for 3.4\,\ensuremath{\upmu\mathrm{m}}, 4.6\,\ensuremath{\upmu\mathrm{m}}, and 12\,\ensuremath{\upmu\mathrm{m}} (for the analysis described in Sect.~\ref{sec:ysos-identification-wise} we do not use the 22\,\ensuremath{\upmu\mathrm{m}} band). Analogously, we estimate completeness limits of 4\,mJy, 3\,mJy and 24\,mJy for 3.4\,\ensuremath{\upmu\mathrm{m}}, 4.6\,\ensuremath{\upmu\mathrm{m}}, and 12\,\ensuremath{\upmu\mathrm{m}}, respectively.

\section{Morphology}
\label{sec:morphology}

Figure~\ref{fig:spitzer_herschel_ausschnitt} shows the far-IR morphology of the Gum\,31 bubble and its environs; the cavity of the H\,II region is clearly delineated. Figure~\ref{fig:wfi_irac_mips} contrasts optical and infrared images of the Gum\,31 bubble. Near the centre of the bubble a cluster of stars is just discernible in the IRAC image. This is NGC\,3324. In the MIPS 24\,\ensuremath{\upmu\mathrm{m}} and PACS 70\,\ensuremath{\upmu\mathrm{m}} images the warm dust surrounding the stars of NGC\,3324 forms an arc-like structure that follows the shape of the inner bubble wall.

A number of pillar-like structures extend from the edge of the bubble into its inner part, especially from its southern rim. Notably, some but not all optical pillars coincide with those seen in the infrared. The photodissociation regions that very sharply delineate the edge of the bubble are well-observable in the IRAC 5.8\,\ensuremath{\upmu\mathrm{m}} image where their fluorescence under the influence of UV radiation can be seen, but also in the MIPS 24\,\ensuremath{\upmu\mathrm{m}} image that shows the emission from the dust grains within. 

\begin{figure*}
\centering
\subfloat[Colour temperature map.]
 {\label{fig:herschel-temp} \includegraphics[height=0.44\textheight]{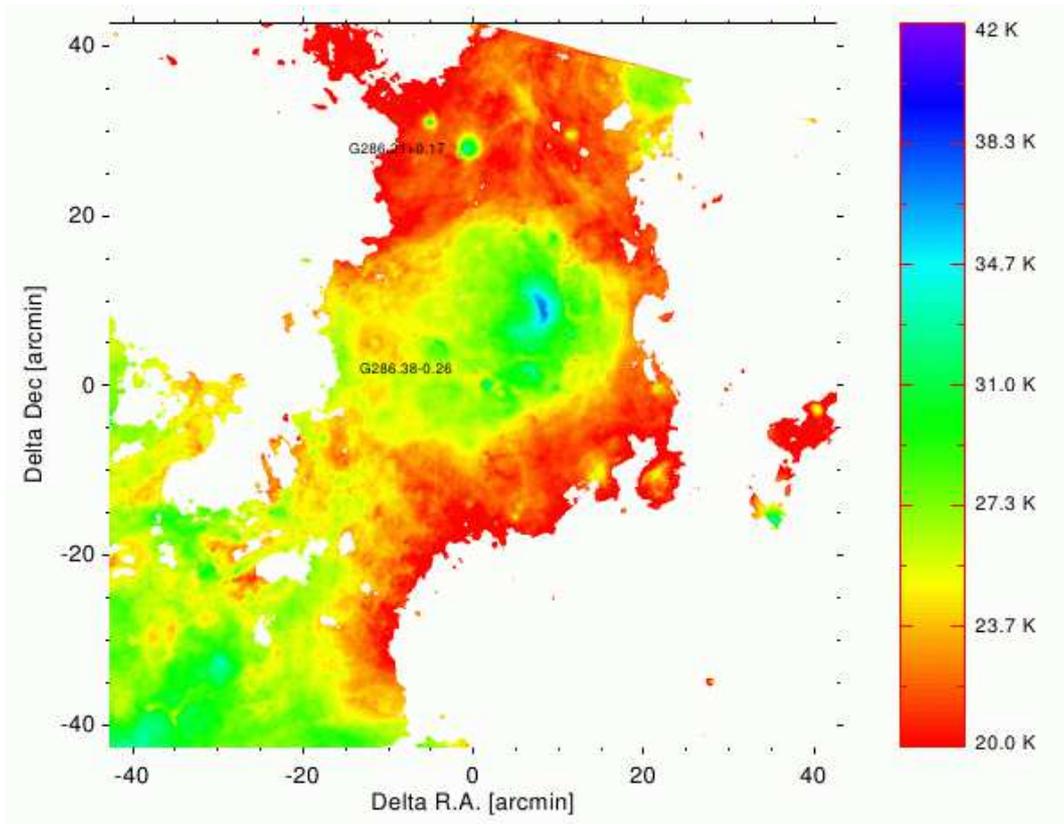}} \\ 
\subfloat[Map of hydrogen column density.]
 {\label{fig:herschel-dens} \includegraphics[height=0.44\textheight]{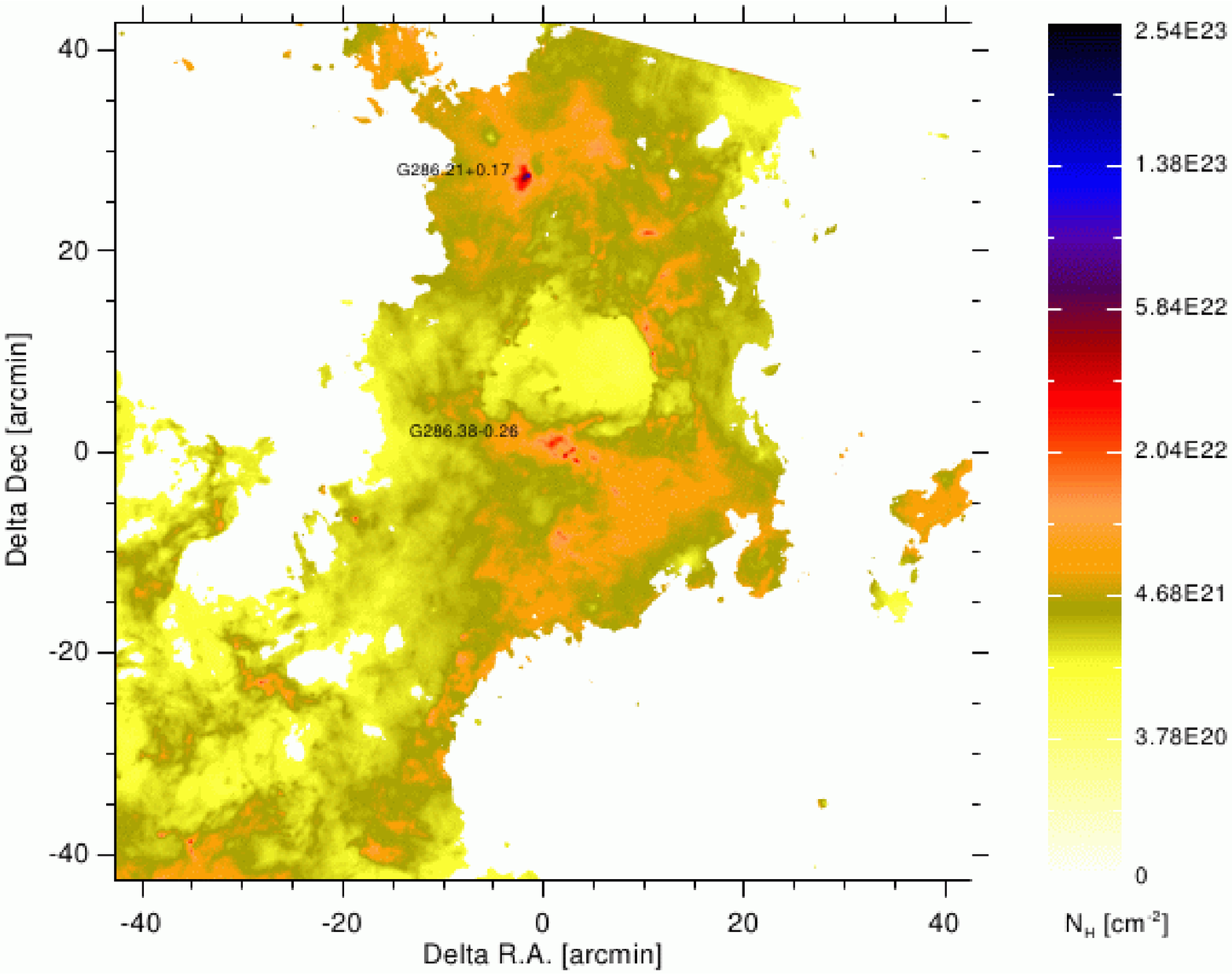}}
  \caption{Maps derived from \textit{Herschel} data. They show the cloud structure in and around the Gum\,31 nebula and the connection to the central parts of the Carina Nebula (just outside the lower left edge of the maps).}
\end{figure*}

Figure~\ref{fig:herschel-temp} shows a colour temperature map for the larger Gum\,31 region, extending downwards into the central Carina Nebula complex. It was derived from the ratio of the PACS 70\,\ensuremath{\upmu\mathrm{m}} and 160\,\ensuremath{\upmu\mathrm{m}} emission as detailed by \citet{preibisch2012}. It shows the temperatures for the H\,II region being comparatively high ($\approx 30$ -- 40\,K), while those of the surrounding clouds are considerably cooler down to 20\,K. The warm dust in NGC\,3324 is clearly seen as an outstandingly blue ($\approx 40$\,K) patch inside the bubble.

The G286.21+0.17 cluster (cf.\ Sect.~\ref{sec:ysos-distribution-byf73}) also stands out. It is about 8\,K warmer than its immediate surroundings. It is also much denser, as can be seen from Fig.~\ref{fig:herschel-dens} which shows the hydrogen column density derived from the colour temperatures as described by \citet{preibisch2012}. We see that the column density inside the bubble is relatively low compared to its surroundings at a few $10^{20}\,\mathrm{cm}^{-2}$ and rises steeply by more than one order of magnitude at the ionisation front. Other than G286.21+0.17, we see some more dense clumps scattered within the bubble rim and beyond it. A notable feature is the cluster G286.38--0.26 (cf.\ Sect.~\ref{sec:ysos-distribution-DBS2003128}) which has a column density of around $3\cdot 10^{22}\,\mathrm{cm}^{-2}$. In sharp contrast to the unusually warm clump G286.21+0.17, in Fig.~\ref{fig:herschel-temp} G286.38--0.26 is seen to be cooler than its surroundings, down to $\approx 20$\,K. The integrated cloud mass derived by \citeauthor{preibisch2012}\ is 186\,700\,\ensuremath{M_\odot}. They derive this value by integrating over the same column density map that we show here, but for a region around Gum\,31 slightly differently defined than the one here.

In the longer IRAC-wavelength emission and especially in the \textit{Herschel} image, a `bridge' of filamentary structure can be seen to extend from Gum\,31 downwards in the direction of the central Carina Nebula, forming a connection. The column density map (Fig.~\ref{fig:herschel-dens}), too, shows that the clouds surrounding Gum\,31 are connected to the clouds in the more central parts of the Carina Nebula. \citet{forte1976} remarks that in the deep optical plates of \citet{lyngaa1972} a filamentary structure connecting the H\,II region H-31 \citep[$\cor \mathrm{Gum\,31}$;][]{hoffleit1953} to NGC\,3372, the central Carina Nebula, can be seen. 

According to \citet{yonekura2005}, the radial velocities of their C\textsuperscript{18}O clumps Nos.~2 -- 6, which surround the Gum\,31 bubble, range from $V_{\rm LSR} = -20.0\,{\rm km\,s^{-1}}$ to $V_{\rm LSR} = -24.1\,{\rm km\,s^{-1}}$. The C\textsuperscript{18}O clumps in the central and northern part of the Carina Nebula (Nos.~8 -- 12) have radial velocities in the $V_{\rm LSR} = -25.8 \ldots -19.9\,{\rm km\,s^{-1}}$ range.
This good agreement of the  radial velocities suggests that the clouds around Gum\,31 and the Carina Nebula are connected and actually part of the same complex. As was argued by \citet{preibisch2012} and in the introduction, we will therefore assume a distance of 2.3\,\ensuremath{\mathrm{kpc}} towards Gum\,31, the same as to the Carina Nebula complex. This number also agrees well with recent distance determinations with independent means (see Sect.~\ref{sec:intro}).

\section{Young stellar objects in the Gum\,31 region}
\label{sec:ysos}

\subsection{Identification of YSO candidates}
\label{sec:ysos-identification}

The selection of cYSOs in the following sections is based on the detection of infrared excesses. It follows, therefore, that only those YSOs exhibiting excess infrared emission can be detected with these methods (cf.\ also Sect.~\ref{sec:ysos-identification-herschel}). Infrared excess emission is indicative of circumstellar material. This means that only YSOs of Class~0 to Class\,II are the subject of this analysis while Class\,III objects must remain undetected.\footnote{YSOs are commonly classified according to the slope of their SEDs \citep{lada1987}, from Class~0 protostars in the main collapse phase through Class\,I YSOs shrouded in envelopes and accreting infalling circumstellar matter and Class\,II YSOs, T~Tauri stars with disks, to Class\,III objects with little or no circumstellar matter.}

\subsubsection{Selection by \textit{Spitzer} colours}
\label{sec:ysos-identification-irac}

\begin{figure}
\centering
\includegraphics[width=0.48\textwidth]{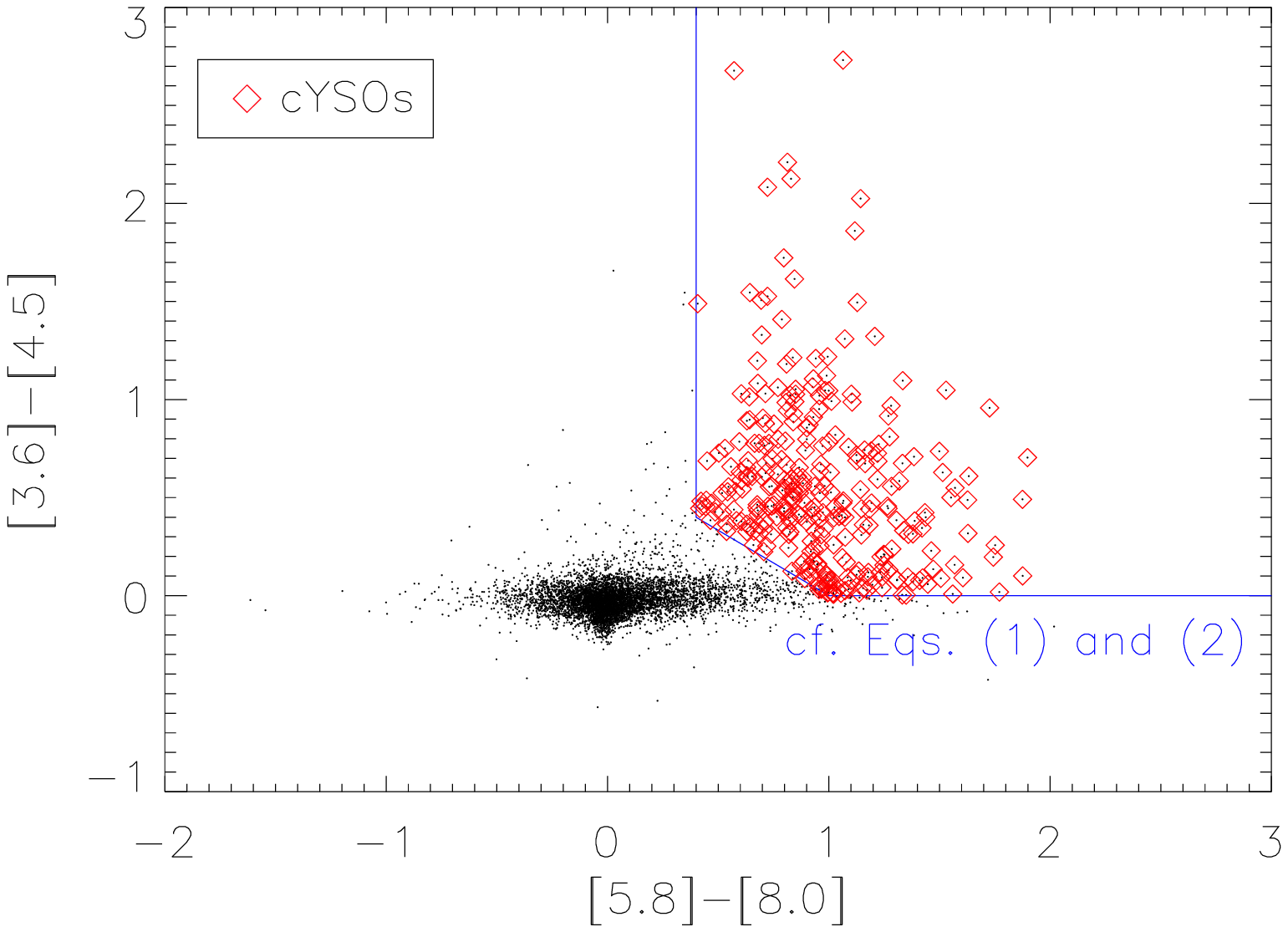}
  \caption{Colour-colour diagram using all \textit{Spitzer} IRAC bands. The criteria adopted for YSO identification follow \citet{fang2012} and \citet{allen2004}.}
  \label{fig:ccdiagram_3645_5880}
\end{figure}

With IRAC data only and following the criteria established for a survey of star-forming region Pismis 24 \citep{fang2012, allen2004} we were able to identify 304  infrared excess sources from plotting the \ensuremath{[3.6]-[4.5]} colour against the \ensuremath{[5.8]-[8.0]} colour for those 6739 point-like sources detected in all four bands.
For identification as a cYSO \citet{fang2012} demand \citep[Eq.~\ref{eq:irac1} follows][]{allen2004}:
\begin{equation}
 [3.6]-[4.5] \geq 0 \quad \mathrm{and} \quad [5.8]-[8.0] \geq 0.4
 \label{eq:irac1}
\end{equation}
\begin{equation}
 [3.6]-[4.5] \geq 0.67 - ([5.8]-[8.0]) \times 0.67 \; .
 \label{eq:irac2}
\end{equation}
The resultant colour-colour diagram is shown in Fig.~\ref{fig:ccdiagram_3645_5880}.
 
If we also take into account the 2MASS magnitudes, we can sample a much larger portion of sources by applying the excess criteria of \citeauthor{winston2007}\ \citep[\citeyear{winston2007}, using \textit{Spitzer} extinction as derived by][]{flaherty2007}. These yield another estimate of the YSOs in the region from the $J-H$ vs.\ $H-[4.5]$ colour-colour diagram. In this way, 2577 cYSOs are identified (from a total sample of 38459 with detections in the three bands).

Within this sample, it is certain that there are contaminants masking as YSOs. AGB stars unfortunately exhibit very similar colours to YSOs and so do background star-forming galaxies. \citet{oliveira2009} in a spectroscopic survey of objects discovered by the \textit{Spitzer} Legacy Program ``From Molecular Cores to Planet-Forming Disks'' (``c2d'') in the Serpens molecular cloud find a contamination of 25\% within their subsample of the brightest objects ($F_{\mathrm{8.0\,\ensuremath{\upmu\mathrm{m}}}} > 3\,\mathrm{mJy}$). They attribute this high proportion to the closeness of the Serpens cloud to the Galactic plane -- something which is also true for Gum\,31 ($b_{\mathrm{Gum\,31}} \approx 0.2\degr$).
We can also estimate the contamination based on the criteria of \citet{winston2007} for the selection of contaminants from the \ensuremath{[5.8]-[8.0]} vs.\ \ensuremath{[4.5]-[5.8]} and \ensuremath{[4.5]-[8.0]} vs.\ \ensuremath{[3.6]-[5.8]} diagrams. This yields an estimated contamination in the IRAC-selected sample of $\lesssim 10\%$.

It is noticeable that the distribution of these cYSOs throughout Gum\,31 and outside it is almost uniform. This is suspicious. We thus conclude that the contamination within this sample due to fore- and background sources is high and do not use this sample for further analysis.

\subsection{Selection by WISE colours}
\label{sec:ysos-identification-wise}

We used WISE catalogue data to search for infrared excess sources. \citet{koenig2012} in their study of massive star-forming regions developed a set of criteria to identify likely YSOs from WISE four-band photometry. We applied these criteria accordingly.

\begin{figure}
\centering
\includegraphics[width=0.48\textwidth]{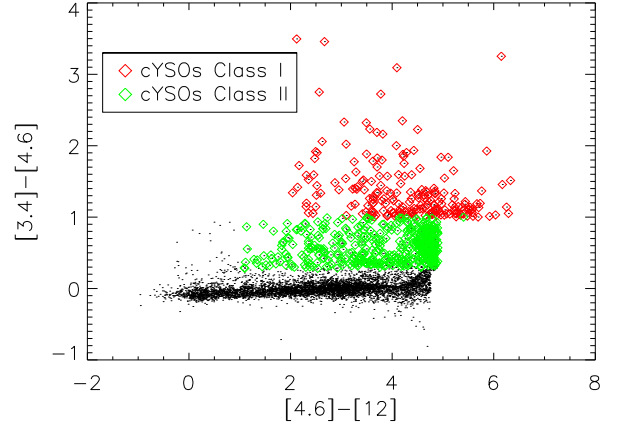}
  \caption{Colour-colour diagram using three out of four WISE bands. The criteria adopted for YSO identification follow \citet{koenig2012}. Probable YSOs are marked in red for Class\,I sources and in green for Class\,II.}
  \label{fig:ccdiagram_3446_4612}
\end{figure}

Using the \ensuremath{[3.4]-[4.6]} and \ensuremath{[4.6]-[12]} colours, partly in combination with the [3.4] and [4.6] magnitudes, probable background objects are removed from the sample before the YSO selection. This includes galaxies (very red in \ensuremath{[4.6]-[12]}), broad-line AGNs (of similar colours as YSOs, but distinctly fainter) and resolved PAH emission regions (redder than the majority of YSOs). From this cleaned sample the IR excess sources are then selected by demanding \citep[Eqs.~\ref{eq:class2} and \ref{eq:class1}:][]{koenig2012}
\begin{equation}
 [3.4]-[4.6] -\sigma_1 > 0.25 \quad \mathrm{and} \quad [4.6]-[12] -\sigma_2 > 1.0 \; ,
 \label{eq:class2}
\end{equation}
where $\sigma$ is the quadratically added uncertainty of the respective magnitudes. Class\,I sources are a subsample of this defined by
\begin{equation}
 [3.4]-[4.6] > 1.0 \quad \mathrm{and} \quad [4.6]-[12] > 2.0
 \label{eq:class1}
\end{equation}
(the rest are Class\,II objects). From this analysis we excluded any data point from the catalogue that did not have a signal-to-noise ratio of 5 or better. Figure~\ref{fig:ccdiagram_3446_4612} shows the \ensuremath{[3.4]-[4.6]} vs.\ \ensuremath{[4.6]-[12]} colour-colour diagram, constructed after probable contaminators had been removed. The total sample before this removal were 10\,128 sources. After removal of probable contaminators, 6669 sources remained. These were plotted in Fig.~\ref{fig:ccdiagram_3446_4612}. Resultant Class\,I sources are marked in red, Class\,II in green. The same colour-coding is used in Fig.~\ref{fig:wise_ysos}, which shows the spatial distribution of the cYSOs. This analysis yields 661 cYSOs of which 207 are Class\,I and 454 are Class\,II.

The distribution of WISE-selected cYSOs follows the cloud structure as expected. \citet{koenig2012} estimate the remaining contamination of the cYSO population selected with their criteria by galaxies to be $\sim 10\deg^{-2}$ which in our $1.2\deg$ field leads to an estimate of $\sim 12$. However, in a region projected onto the Galactic disk as the Carina Nebula we expect a high level of contamination mainly from background and foreground stars. To estimate the total contamination we performed the same analysis as described above for two circular areas of 30\arcmin\ radius well outside the Carina Nebula (i.\,e.\ regions that we expect to be relatively free of YSOs), one centred around $\upalpha_{2000} = 10$:51:52.5, $\updelta_{2000} = -59$:14:30, the other around $\upalpha_{2000} = 10$:25:10.7, $\updelta_{2000} = -58$:52:00. They were chosen as having a similar proximity to the Galactic plane as Gum\,31 ($b=0.15$ and $b=-1.18$; Gum\,31: $b=-0.17$). Furthermore, we carefully selected fields that were as free of CO and H$\upalpha$ emission as possible and appeared as empty as possible in the IRAS 100\,\ensuremath{\upmu\mathrm{m}} images.

In those control fields we find a mean cYSO density of $97\,\mathrm{deg}^{-2}$; with the cYSOs spread homogeneously. In the Gum\,31 region we determine a cYSO density of $550\,\mathrm{deg}^{-2}$. This leads us to expect a contamination of around 18\% ($\sim 120$ cYSOs) for the Gum\,31 cYSO sample. For the IRAC data we were not able to conduct a similar comparison as we do not have photometry data for the full study area around Gum\,31 and none for regions outside the central nebula.

We therefore deem the WISE-selected sample to be more reliable and base our conclusion in the following sections predominantly on this sample, although we will occasionally describe a classification with IRAC. In general, though, those should be treated as less reliable.

\subsection{Identification of protostars from \textit{Herschel} data}
\label{sec:ysos-identification-herschel}

With the methods described in Sect.~\ref{sec:data-herschel} and in more detail by \citet{gaczkowski2012}, we obtained a point-source catalogue for the \textit{Herschel} data. 
Although we consider only `point-like' \textit{Herschel} sources in the following, it is important to keep in mind that the relatively large PSF corresponds to quite large physical scales at the 2.3\,\ensuremath{\mathrm{kpc}} distance of Gum\,31. In the PACS 70\,\ensuremath{\upmu\mathrm{m}} map, all objects with an angular [spatial] extent of up to $\approx 5\arcsec$ [0.056\,\ensuremath{\mathrm{pc}}] are compact enough to appear `point-like'. For the SPIRE 250\,\ensuremath{\upmu\mathrm{m}} map, this is true for sources of up to $\approx 18\arcsec$ [0.20\,\ensuremath{\mathrm{pc}}]. This shows immediately that (pre-stellar) cloud cores, which have typical radii of $\la 0.1$\,\ensuremath{\mathrm{pc}}, cannot be (well) resolved in the \textit{Herschel} maps and may appear as compact `point-like' sources.
This implies that YSOs in all their evolutionary stages can, in principle, appear as point-like sources in our \textit{Herschel} maps. However, the possibility to detect an object in a specific stage depends strongly on its properties: As described by \citet{gaczkowski2012}, many pre-stellar cores ($\gtrsim 1$ -- 2\,\ensuremath{M_\odot}) and embedded protostars ($\gtrsim 1$\,\ensuremath{M_\odot}) will be easily detectable, while most of the more evolved pre-main sequence stars with disks should remain undetected. In this last case the detection limit depends on the disk mass; as shown by \citeauthor{gaczkowski2012}\ a 1\,\ensuremath{M_\odot} YSO is still detectable if it has a disk mass of $\gtrsim 0.5$\,\ensuremath{M_\odot}, this sinks to $\gtrsim 0.1$\,\ensuremath{M_\odot} for a 3\,\ensuremath{M_\odot} YSO and $\gtrsim 0.01$\,\ensuremath{M_\odot} for a 6\,\ensuremath{M_\odot} YSO.

\begin{figure*}
\sidecaption
\includegraphics[width=12cm]{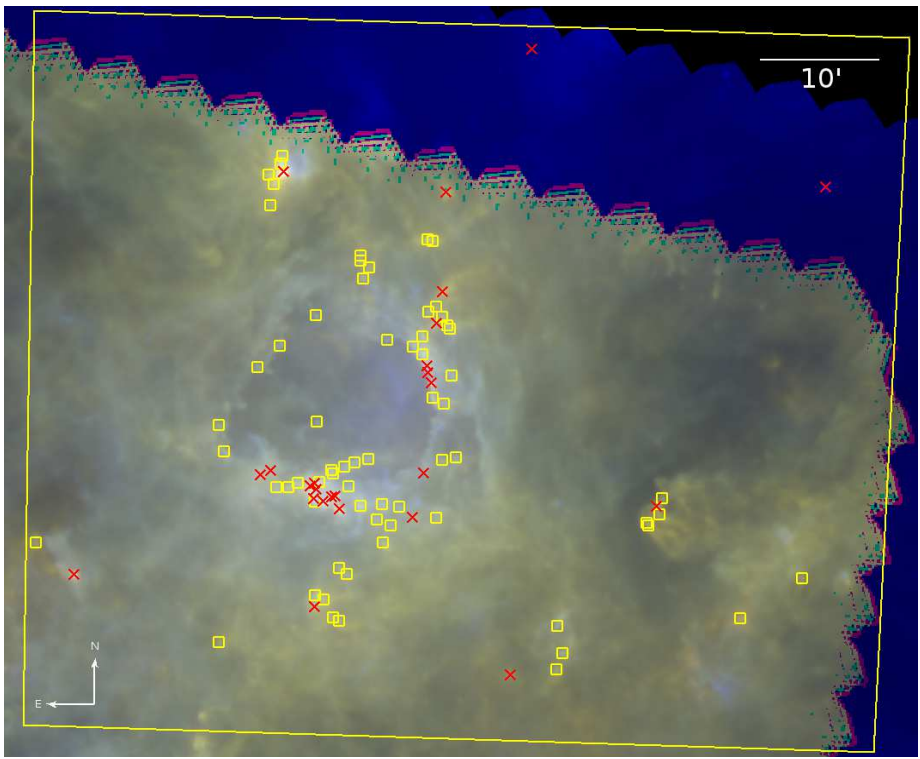}
  \caption{\textit{Herschel} RGB image (red: 500\,\ensuremath{\upmu\mathrm{m}}, green: 250\,\ensuremath{\upmu\mathrm{m}}, blue: 70\,\ensuremath{\upmu\mathrm{m}}) with positions of all \textit{Herschel} point-like sources detected in at least two bands overlaid. Red crosses show the position of protostellar, yellow boxes those of prestellar cores. The large yellow rectangle marks the borders of the region analysed.}
  \label{fig:all_herschel}
\end{figure*}

This means that while with IRAC we find mainly Class\,II YSOs and a number of Class\,I YSOs, with \textit{Herschel} the emphasis is on Class~0 protostars, with some Class\,I stars. This also implies that the overlap between both is relatively small (cf.\ Sect.~\ref{sec:herschelspitzer}). \citet{raganEtal2012} presented radiative transfer models of starless cores and protostellar cores and investigated the detectability of these classes of objects. Their models showed that the SEDs of starless (i.\,e.\ pre-stellar) cores typically peak around 200 --300\,\ensuremath{\upmu\mathrm{m}} and drop very steeply towards shorter wavelengths. Their model fluxes at 70\,\ensuremath{\upmu\mathrm{m}} (scaled to the
distance of the CNC) are several orders of magnitudes below our detection limits. Protostellar cores, on the other hand, have much stronger fluxes at PACS wavelengths. Guided by these results, we can take a detection at 70\,\ensuremath{\upmu\mathrm{m}} as an indication for the protostellar nature of the source, whereas \textit{Herschel} sources without 70\,\ensuremath{\upmu\mathrm{m}} detection would then be pre-stellar cores. In Fig.~\ref{fig:all_herschel} these two classes are indicated separately.

\citet{gaczkowski2012} argue that in a sample of \textit{Herschel} point-like objects in the Carina Nebula it is very unlikely for contamination by evolved stars or extragalactic objects to occur. The same reasoning applies to Gum\,31.
The photometric data for those sources that fall within the Gum\,31 area can be found in the paper of \citet{gaczkowski2012}.

\subsection{Spatial distribution of the cYSOs}
\label{sec:ysos-distribution}

It is notable that both Class\,I and Class\,II sources are found predominantly within the interior of the Gum\,31 bubble or along its rim. They tend to occur in clusters and in their distribution are often correlated with the \citet{yonekura2005} molecular cloud clumps (see below).
When in the following we refer to a cYSO with a number preceded by `[CNA2008]' this source was classified as a cYSO by \citet{cappa2008}.

\begin{figure*}
\centering
  \includegraphics[width=\textwidth]{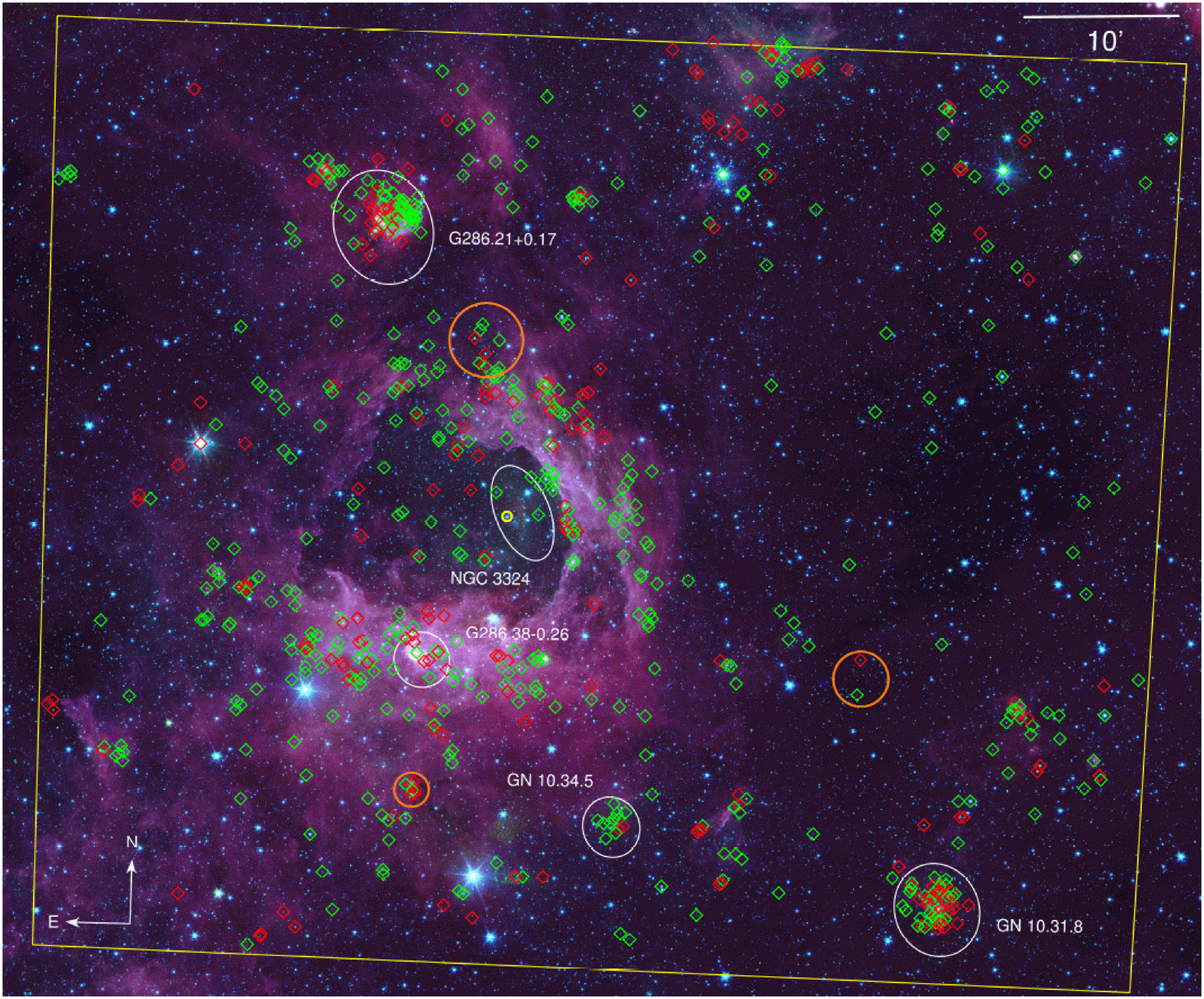}
  \caption{\textit{Spitzer} IRAC RGB image (red: 8.0\,\ensuremath{\upmu\mathrm{m}}, green: 4.5\,\ensuremath{\upmu\mathrm{m}}, blue: 3.6\,\ensuremath{\upmu\mathrm{m}}) with positions of WISE-identified cYSOs overlaid. Class\,I sources are marked by red diamonds, Class\,II in green. The white ellipses represent the clusters discussed in Sect.~\ref{sec:ysos-distribution}, with the yellow circle marking the multiple star HD\,92206 which contains the O stars in NGC\,3324. The orange circles mark the cYSO clusters discussed in Sect.~\ref{sec:ysos-distribution-undescribed}. The yellow rectangle marks the borders of the region analysed.}
  \label{fig:wise_ysos}
\end{figure*}

\subsubsection{NGC\,3324}
\label{sec:ysos-distribution-ngc3324}

The cluster NGC\,3324 appears prominent in the \textit{Spitzer} images, containing $\approx 200$ point-like sources. 
Within the cluster itself, we identify only a single cYSO with WISE, J103706.9--583710 at the western edge of the cluster (Class\,II). From the \ensuremath{[3.6]-[4.5]} vs.\ \ensuremath{[5.8]-[8.0]} diagram (IRAC) we identify half a dozen cYSOs coincident with the cluster NGC\,3324 and distributed about 40\arcsec\ towards the north-east of the cluster centre. The very small fraction of stars with detectable infrared excess ($\approx$ 0.5\%) supports previous age estimates of $\gtrsim 3$\,\ensuremath{\mathrm{Myr}} for this cluster.

\subsubsection{cYSOs in the rim of the bubble}
\label{sec:ysos-distribution-bubblerim}

Numerous cYSOs are found lined up along the ionisation front of the Gum\,31 bubble to the west of NGC\,3324. A dozen cYSOs are found right behind the edge of the ionisation front. Among them are J103653.9--583719, J103653.3--583754, and J103652.4--583809, three Class\,I candidates that are found along the ridge traced in IRAC and \textit{Herschel} images at the very edge of the bubble, neighbouring NGC\,3324. Two of them, J103653.9--583719 and J103652.4--583809, will be discussed in Sect.~\ref{sec:hhjets} as possible sources of Herbig-Haro jets.

Behind this `first row' of Class\,I candidates, there is a `second row' of five Class\,II candidates, all lined up about 19\arcsec\ behind the ionisation front. Further northwards along the rim there are four more Class\,II candidates, one similarly behind the front and three right along it. 

In Fig.~\ref{fig:wise_ysos} it is evident that a major part of the cYSOs is located along the bubble edges, similar to what is observed in comparable bubbles associated with H\,II regions \citep[e.\,g.][]{dewangan2012}. Their distribution follows its circular shape and their number sharply goes down outside the cloud structure that appears magenta in the image. This is suggestive of triggered star-formation in a `collect and collapse' scenario as described by \citet{whitworth1994}.

Following analytical models for this scenario, a stellar wind bubble produced by three massive stars with spectral types O6.5, O6.5 and O9.5 could reach a radius of approximately 9 -- 11\,\ensuremath{\mathrm{pc}} in 1.5 -- 2.0\,\ensuremath{\mathrm{Myr}} assuming initial cloud densities in a range of 500 -- $1000\,\ensuremath{\mathrm{cm}}^{-3}$. The radius at which fragmentation occurred would then vary between 7\,\ensuremath{\mathrm{pc}} and 11\,\ensuremath{\mathrm{pc}}. To estimate the mechanical luminosities emitted by the three most massive stars in NGC\,3324, we used values from \citet{smith2006} (and the erratum \citep{smith2006_erratum}) for stars with the same spectral types (Luminosity class V) observed in the Carina Nebula. The $\sim 15\arcmin$ diameter of the Gum\,31 H\,II region corresponds to 10\,\ensuremath{\mathrm{pc}} at a distance of 2.3\,\ensuremath{\mathrm{kpc}} which agrees very well with the values derived from the model. The fact that we also find numerous embedded cYSOs in and near the rim of the bubble is consistent with star-formation according to the `collect and collapse' model.

\subsubsection{cYSOs in pillars}
\label{sec:ysos-distribution-pillars}

\begin{figure*}
\centering
    \includegraphics[width=\textwidth]{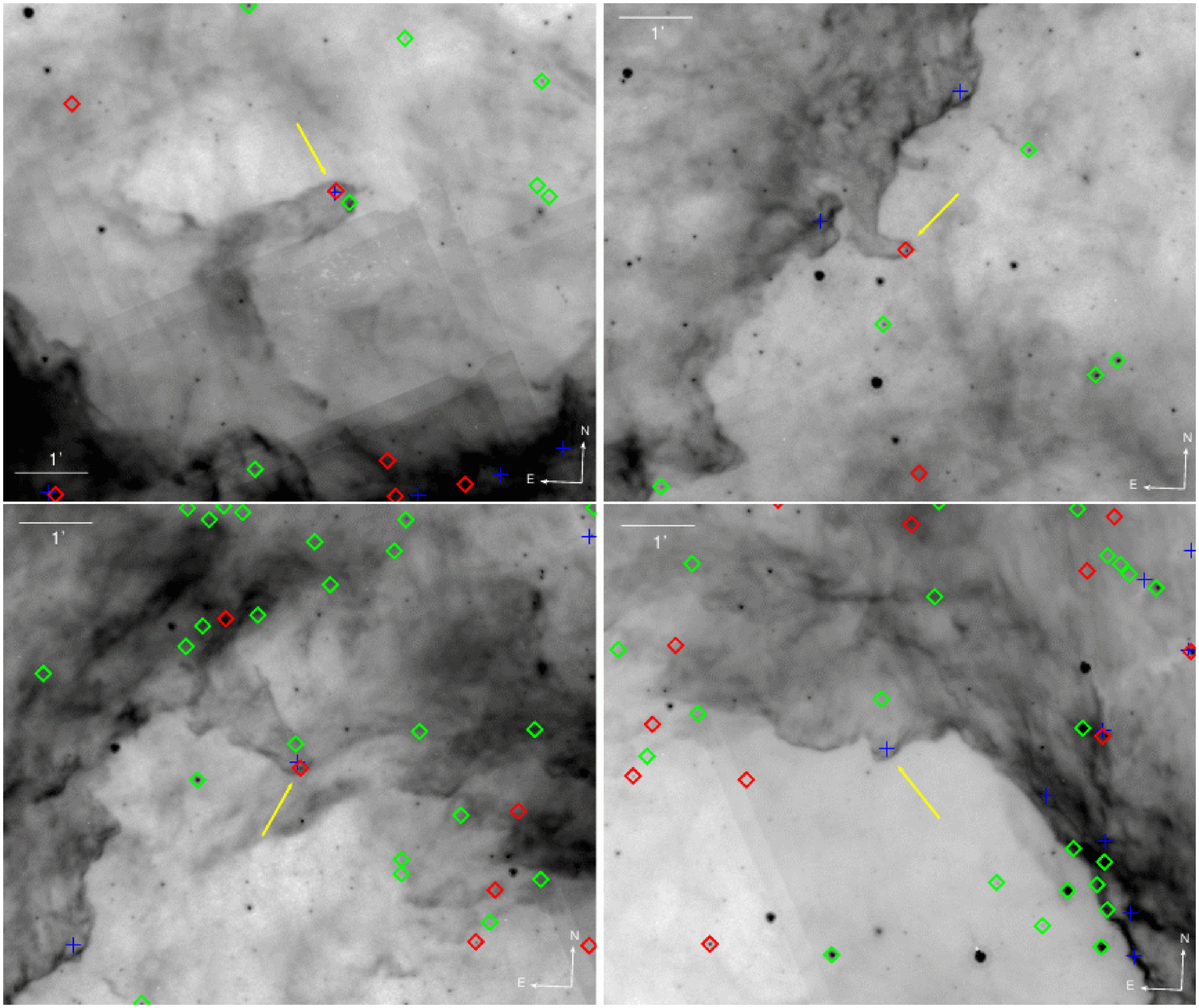}
  \caption{In a \textit{Spitzer} IRAC 8.0\,\ensuremath{\upmu\mathrm{m}} image, the yellow arrows mark the positions of four remarkable cYSOs found in the very tips of pillars. 
  Positions of WISE-identified cYSOs are overlaid as in Fig.~\ref{fig:wise_ysos}: Class\,I sources are marked by red diamonds, Class\,II sources in green. Blue crosses mark \textit{Herschel} point-like sources detected in at least two bands.}
  \label{fig:spitzer-herschel_266}
\end{figure*}

In the southern and eastern part of the bubble edge we find a number of small pillars extending into the bubble interior. They can be seen in the optical and infrared in Fig.~\ref{fig:wfi_irac_mips} and a closeup of the IR pillars is shown in Fig.~\ref{fig:spitzer-herschel_266}. Within four of them we find cYSOs in their very tips, reminiscent of what is observed in the central Carina Nebula (e.\,g.\ the South Pillars region). One of them, J103806.6--584002, coincides with a \textit{Herschel} point-like source (cf.\ Sect.~\ref{sec:herschelspitzer-seds}) and is therefore most probably a protostar. This suggests that radiative triggering is at work, very similar to the processes seen in the South Pillars \citep{smith2010_spitzer, gaczkowski2012}.

\subsubsection{G286.21+0.17}
\label{sec:ysos-distribution-byf73}

The cloud clump G286.21+0.17 ([DBS2003]\,127, BYF73) is located about 12\arcmin\ north of the rim of the Gum\,31 bubble. The structure of this clump was recently studied in several molecular lines by \citet{barnes2010}. They determined a diameter of $\sim 0.9$\,\ensuremath{\mathrm{pc}}, a luminosity of $\sim 2$ -- $3 \cdot 10^4\,\ensuremath{L_\odot}$, and estimated a clump mass of about 20\,000\,\ensuremath{M_\odot} (a value that is about forty times larger than the previous mass estimate based on millimetre data by \citeauthor{faundez2004}\ \citeyear{faundez2004}). Their derived mass infall rate of $\sim 3.4 \cdot 10^{-2}\,\ensuremath{M_\odot}\,{\rm yr}^{-1}$ would be the highest mass infall rate yet seen, if confirmed.
These properties make this cloud a particularly interesting site of possible massive-star formation.

In our \textit{Herschel} maps, this clump appears as a very bright and prominent compact feature. In Fig.~\ref{fig:byf73}, we compare its morphology in the \textit{Herschel} far-IR bands to a near-IR image from our VISTA data and a \textit{Spitzer} mid-IR image. As already discussed by \citet{barnes2010}, a young stellar cluster, surrounded by diffuse nebulosity, is located immediately north-west of the clump. This cluster appears very prominent in Fig.~\ref{fig:wise_ysos} as well and contains $\sim 45$ cYSOs. In the centre of the clump itself, the \textit{Spitzer} images show two bright point sources with an angular separation of $7.6\arcsec$ (Fig.~\ref{fig:byf73}).
They can be identified with the 10\,\ensuremath{\upmu\mathrm{m}} point sources J103832.08--581908.9 and J103832.71--581914.8 that were detected as counterparts of the MSX/RMS massive cYSO G286.2086+00.1694 described by \citet{mottram2007}, for which a bolometric luminosity of 7750\,\ensuremath{L_\odot} was determined.

\begin{figure}
\centering
\includegraphics[width=0.48\textwidth]{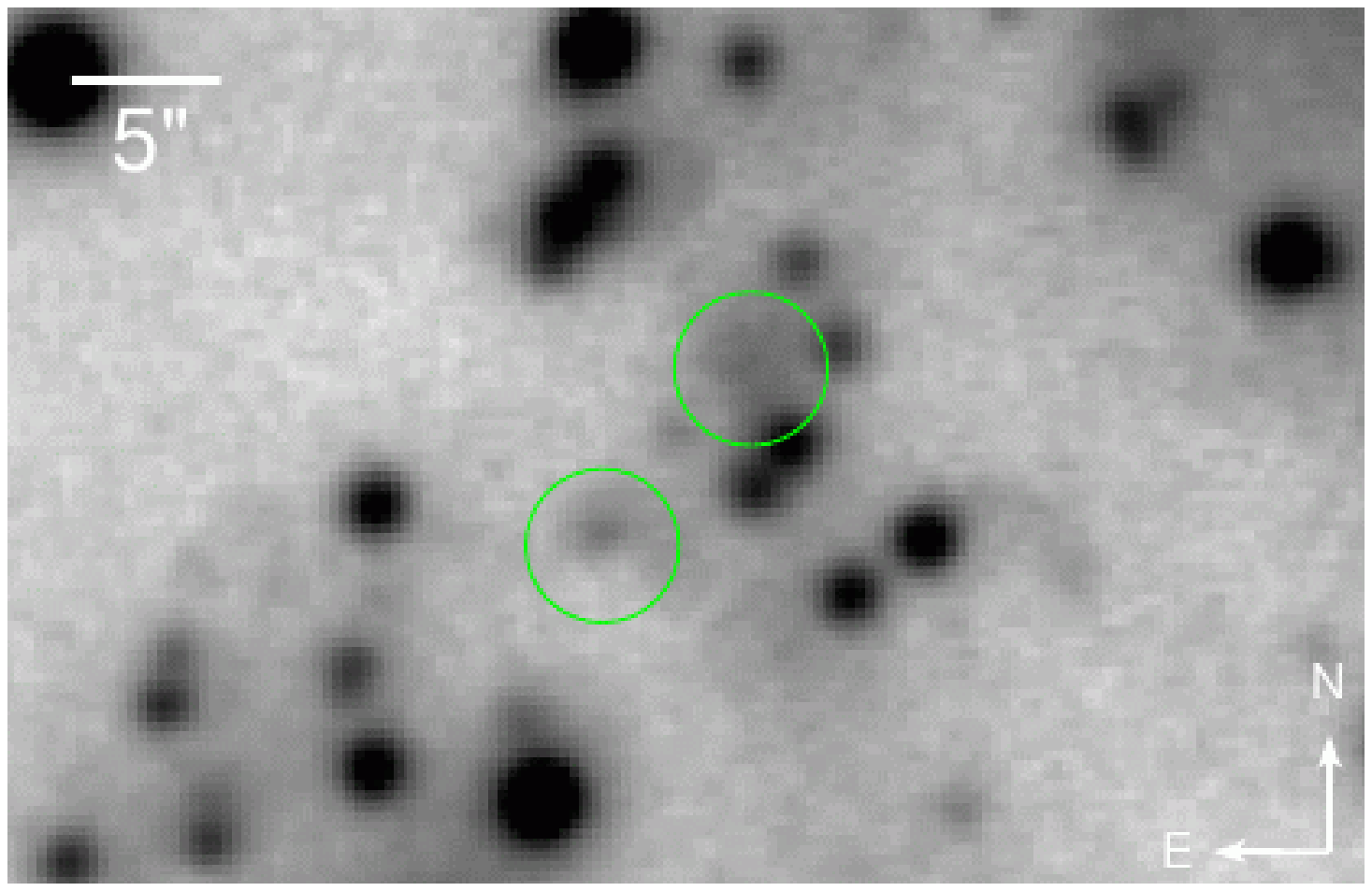}
  \caption{Close-up of the VISTA \emph{H}-band image (Fig.~\ref{fig:byf73}) around the two bright IRAC sources (green circles) within the cluster G286.21+0.17.}
  \label{fig:byf73_zoom}
\end{figure}

\begin{figure*}
\centering
\includegraphics[width=\textwidth]{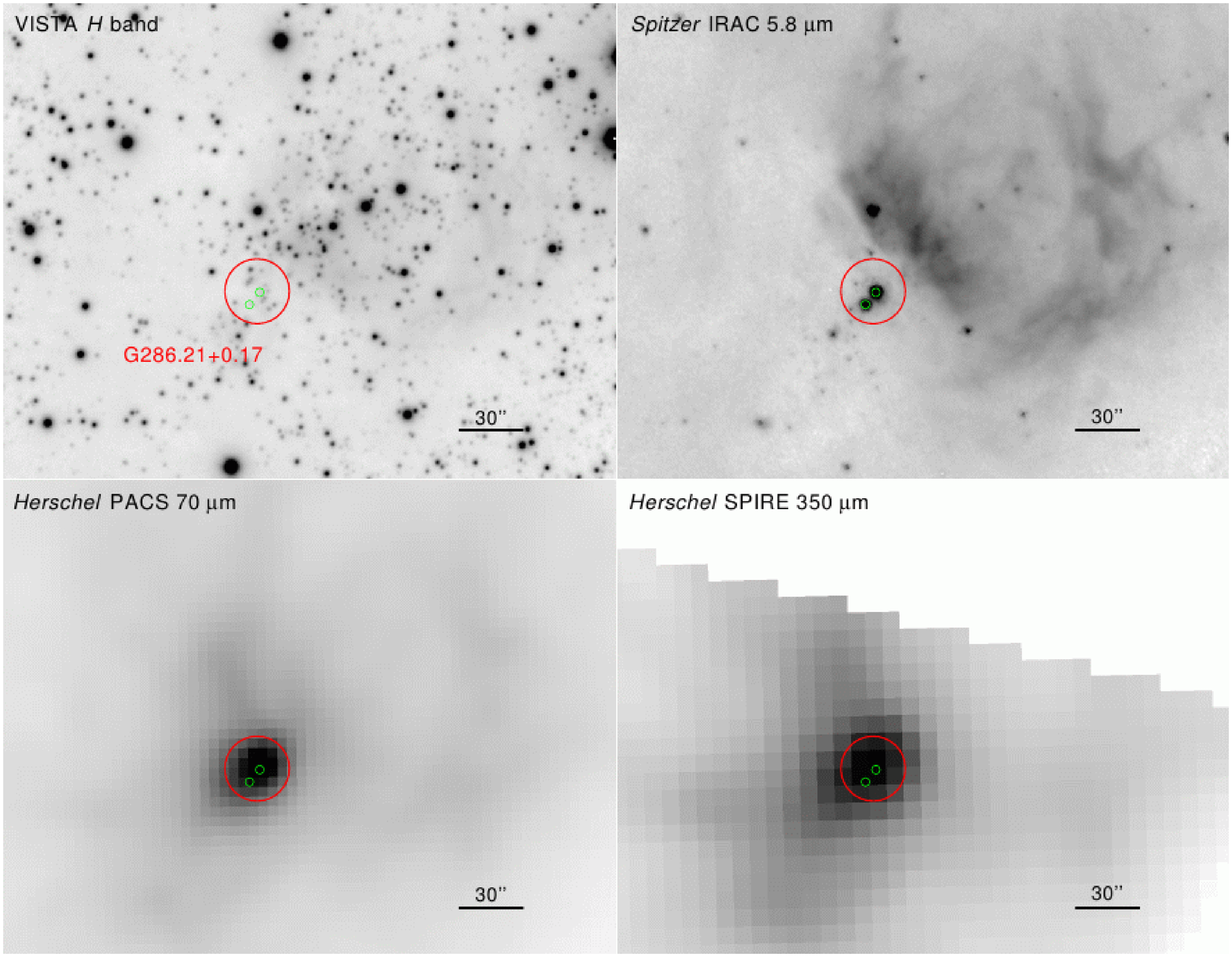}
  \caption{Cluster G286.21+0.17 (red circle; cf.\ Sect.~\ref{sec:ysos-distribution-byf73}) and its immediate surroundings from near to far-IR. The two green circles mark the two bright \textit{Spitzer}-resolved sources within it. (A close-up of the VISTA \emph{H}-band image is shown in Fig.~\ref{fig:byf73_zoom}.)}
  \label{fig:byf73}
\end{figure*}

The VISTA \emph{H}-band image shows very faint diffuse nebulosities at the location of these two mid-IR sources (see the close-up in Fig.~\ref{fig:byf73_zoom}). This is consistent with the idea that these two objects represent deeply embedded YSOs in the protostellar evolutionary phase. With WISE the two sources are not resolved but run into one that we classify as a Class\,I candidate.

The peak of the \textit{Herschel} PACS 70\,\ensuremath{\upmu\mathrm{m}} emission is centred on the mid-IR source J103832.0--581908. A two-dimensional Gaussian fit to the 70\,\ensuremath{\upmu\mathrm{m}} emission yields a FWHM size of $13.6\arcsec \times 11.3\arcsec$, which is clearly larger than the FWHM size of $10\arcsec \times 10\arcsec$ measured for several isolated point-like sources in the same map. With an angular distance of $7.6\arcsec$, the emission of the two mid-IR sources cannot be resolved in the \textit{Herschel} maps, but the measured direction of the elongation towards J103832.71--581914.8 suggests that both contribute to the observed far-IR emission.

A second cYSO is found in the south-east of the clump, at the projection of the line connecting J103832.08--581908.9 and J103832.71--581914.8. It probably corresponds to the IRAC-identified source at $\upalpha_{2000} = 10$:38:33.6, $\updelta_{2000} = -58$:19:22. There is a third identified cYSO, situated at the very edge of the clump, at $\upalpha_{2000} = 10$:38:35.0, $\updelta_{2000} = -58$:18:44. These three are the only cYSOs identified in the cluster with WISE, while in the immediately adjoining cluster of stars visible in the IRAC images a large number of cYSOs is found. There are $\sim 45$ overall, with a slight majority of Class\,II sources over Class\,I sources. If there is any trend in their spatial distribution, Class\,I are found with slight emphasis to the south-east, while Class\,II sources tend to be located towards the north-west. 

Our colour-temperature map (Fig.~\ref{fig:herschel-temp}), constructed from the \textit{Herschel} 70\,\ensuremath{\upmu\mathrm{m}} and 160\,\ensuremath{\upmu\mathrm{m}} maps, shows that the cloud temperatures range from $\la 20$\,K at the edge of the clump to $\sim 25$ -- 30\,K in the clump centre, and up to 33\,K in the nebulosity surrounding the stellar cluster north-west of the clump.

In our \textit{Herschel} column-density map (Fig.~\ref{fig:herschel-dens}), the level $N_{\rm H} = 2 \cdot 10^{22}\,\mathrm{cm}^{-2}$ traces the shape of the clump. This agrees very well with the morphology as seen in the 1.2\,mm map shown by \citet{faundez2004}. The peak value of the column density is found to be $ 1.4 \cdot 10^{23}\,\mathrm{cm}^{-2}$ and corresponds to a visual
extinction of $A_V \approx 70$\,mag. From our column density map we determined the mass of the clump by integrating over a $200\arcsec \times  200\arcsec$  ($2.23\,\mathrm{pc} \times 2.23$\,\ensuremath{\mathrm{pc}}) box around the clump and subtracting the local background level. This yields a clump mass of 2\,105\,\ensuremath{M_\odot}. This value is nearly five times as large as the 470\,\ensuremath{M_\odot} derived by \citet{faundez2004}, but a factor of ten smaller than the 20\,000\,\ensuremath{M_\odot} estimated by \citet{barnes2010}. This seems to suggest that the mass infall rate estimated by \citet{barnes2010} is also too high, and this clump is not as extreme as suspected.

Its far-IR fluxes as derived from our \textit{Herschel} data using elliptical apertures for photometry with GAIA\footnote{\url{http://astro.dur.ac.uk/~pdraper/gaia/gaia.html}} are 1661\,Jy, 2261\,Jy, 1293\,Jy, 653\,Jy and 297\,Jy for 70\,\ensuremath{\upmu\mathrm{m}}, 160\,\ensuremath{\upmu\mathrm{m}}, 250\,\ensuremath{\upmu\mathrm{m}}, 350\,\ensuremath{\upmu\mathrm{m}} and 500\,\ensuremath{\upmu\mathrm{m}}, respectively, which results in an integrated far-IR luminosity (70\,\ensuremath{\upmu\mathrm{m}} to 1.3\,mm) of $L_{\mathrm{int}} \approx 9000\,\ensuremath{L_\odot}$.

This is clearly one of the most luminous clumps in the CNC. Its mass is rather high, but probably not as high as previously suggested. It may form stars with $\la 10\,\ensuremath{M_\odot}$, but is probably not massive enough for the formation of high-mass stars with $M_\ast \ga 20\,\ensuremath{M_\odot}$.

\subsubsection{G286.38--0.26}
\label{sec:ysos-distribution-DBS2003128}

\begin{figure*}
\centering
\includegraphics[width=\textwidth]{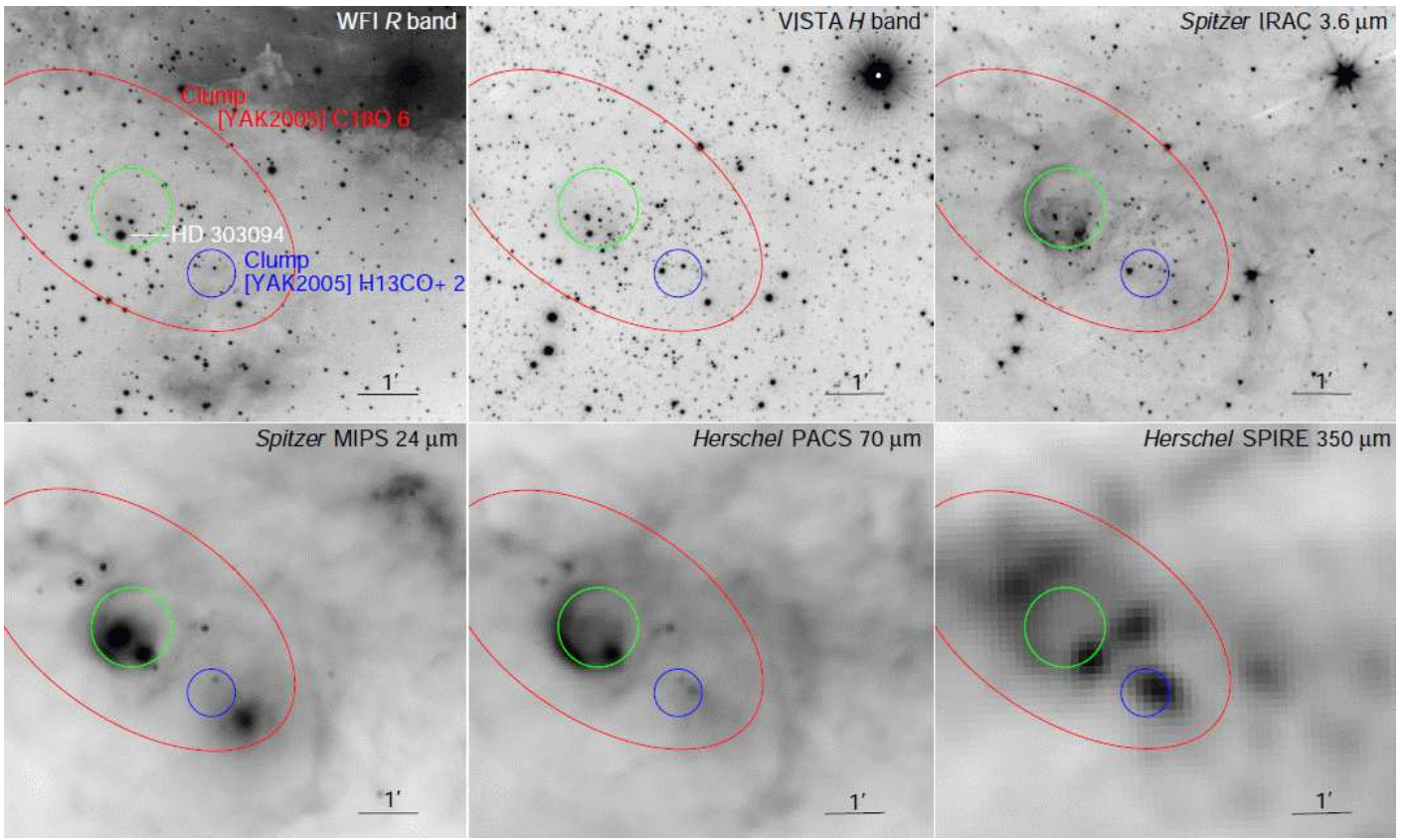}
  \caption{Cluster G286.38--0.26 from near- to far-infrared. The large red ellipse marks the approximate outline of the C\textsuperscript{18}O clump No.~6 \citep[{[}YAK2005{]} C18O 6;][]{yonekura2005}. The blue circle marks H\textsuperscript{13}CO\textsuperscript{+} clump No.~2 ([YAK2005] H13CO+ 2), the green circle indicates the shape of the arc-like nebulosity visible in the \textit{Spitzer} IRAC images.}
  \label{fig:dbs3003_128}
\end{figure*}

The \textit{Spitzer} IRAC images show a prominent dense cluster of several dozen stars at the southern edge of the Gum\,31 bubble, which is surrounded by bright diffuse nebulous emission (Fig.~\ref{fig:dbs3003_128}). This cluster is listed as [DBS2003]\,128 by \citet{dutra2003}. It is spatially coincident with the extended ($r = 2$\,\ensuremath{\mathrm{pc}}) C\textsuperscript{18}O clump\footnote{We note that \citet{yonekura2005} denoted these structures as `cores'; however, according to the definition that \emph{cores} are very compact clouds (with typical sizes of $\sim 0.1$\,\ensuremath{\mathrm{pc}} or less), out of which individual stellar systems form, these clouds are better characterised as \emph{clumps} (i.\,e.\ relatively large dense clouds linked to the formation of small stellar clusters).} No.~6 \citep{yonekura2005}.

The nebulosity around the stellar cluster displays a remarkable arc-like shape at the eastern edge. Projecting it into a full circle, it would have around 42\arcsec\ diameter in IRAC images and 62\arcsec\ in \textit{Herschel} images. The centre of this circle would be around $\upalpha_{2000} = 10$:38:03, $\updelta_{2000} = -58$:46:19. The star J10380461--5846233, cYSO [CNA2008]\,21, is found close ($\sim 11\arcsec$) to this central position of the arc.
In the \textit{Spitzer} bands, the cYSO shows strongly increasing brightness with wavelength. In the MIPS 24\,\ensuremath{\upmu\mathrm{m}} image, it is the brightest point-source in the cluster. It was detected as a mid-IR source by MSX and is listed as G286.3773-00.2563 in the MSX6C catalogue. With WISE data we classify it as a Class\,II cYSO.
The star is not detected in any of our \textit{Herschel} far-IR images. Using USNO-B optical catalogue data, 2MASS, IRAC, MIPS, and WISE photometry and \textit{Herschel} upper limits we employed the online SED fitter by \citet{robitaille2007} to construct an SED and thus estimate the (proto-) stellar parameters. The stellar mass is estimated to be $\approx 5.8\,\ensuremath{M_\odot}$ for the best-fit model, the luminosity $\approx 238\,\ensuremath{L_\odot}$.
Within the arc three further WISE Class\,I cYSOs are seen, J103805.8--584542, J103758.4--584648 and J103800.7--584654.

The optically brightest star in the cluster is HD\,303094, for which a spectral type A2 is given by \citet{nesterov1995}. It is located about 17\arcsec\ south of J10380461--5846233 and the centre of the arc.
According to the \citet{pickles2010} survey of all-sky spectrally matched Tycho-2 stars it may be a foreground star (distance: 886\,\ensuremath{\mathrm{pc}}).

Strong far-infrared emission from the region of this cluster was detected with IRAS (point source IRAS\,10361--5830).
Our \textit{Herschel} maps resolved this far-IR emission into ten point-like sources in the area of the clump. The only \textit{Herschel} source with near-IR counterpart is J103801.4--584641. The IRAC image at this point is dominated by strong nebular emission.
There is extended emission in a confined region a few arcseconds east of the \textit{Herschel}-identified point-like source, but since it is not well-resolved and the identification with the \textit{Herschel}-identified source is not unambiguous, we do not include it in the sample studied in Sect.~\ref{sec:herschelspitzer}.
The two \textit{Herschel}-identified point-like sources north-east of the arc-like nebula that are also detected as bright sources in the MIPS maps are J103810.2--584527 and J103807.2--584511, both have no clear near-IR counterparts.
A similar case is J103754.0--584614, which has a very faint nebulous near-IR counterpart in the VISTA image. All three are also detected as point-like sources in our IRAC images. They are included in Table~\ref{tab:fluxes} (online material), but only for J103807.2--584511 we obtained the model parameters listed in Table~\ref{tab:modell} as for the others the quality of the SED fit was not sufficient. Additionally, all three sources described are classified as Class\,I sources from WISE data.

In the south-western part of this extended C\textsuperscript{18}O clump, \citet{yonekura2005} detected a compact ($r = 0.27$\,\ensuremath{\mathrm{pc}}) H\textsuperscript{13}CO\textsuperscript{+} clump (their clump No.~2); with a central density of $n(\mathrm{H}_2) = 6.8 \cdot 10^4 \, \mathrm{cm}^{-3}$ this is the densest of all the clumps they detected in their survey of the Carina Nebula complex. The \textit{Herschel}-identified point-like source J103750.8--584718 coincides with this clump. With IRAC, there are several small point-like sources seen to be coincident with it and an identification is therefore impossible. An IRAC-detected source slightly to the south-west of it, J103749.3--584722, is identified as a Class\,I WISE cYSO.

This area also contains the A0 supergiant HD\,92207. This star has a strong near-IR excess, was detected as a 12\,\ensuremath{\upmu\mathrm{m}} source with IRAS (IRAS\,10355--5828), and is seen as a point-like source surrounded by nebulosity in the MIPS 24\,\ensuremath{\upmu\mathrm{m}} image. Neither the star not the surrounding nebulosity can be seen in the \textit{Herschel} images.

\subsubsection{Other structure}
\label{sec:ysos-distribution-undescribed}

Towards the south-west of the Gum\,31 shell, two reflection nebulae are found, GN\,10.34.5 and GN\,10.31.8 (both are marked with white ellipses in Fig.~\ref{fig:wise_ysos}). Both are very clearly delineated by cYSOs and coincide with two of the most conspicuous clusters in our field of view. GN\,10.34.5 is visible in the \textit{Spitzer} RGB image (especially the 4.5\,\ensuremath{\upmu\mathrm{m}} band) and seen in projection with a dozen cYSOs. GN\,10.31.8, on the other hand, is both larger in angular extent and coincident with four times the number of cYSOs. It, too, is very conspicuous in the 4.5\,\ensuremath{\upmu\mathrm{m}} band.

There also is a smaller cluster of about two dozen IRAC point-like sources centred at $\upalpha_{2000} = 10$:38:03, $\updelta_{2000} = -58$:55:09 within the Gum\,31 shell (orange circle in Fig.~\ref{fig:wise_ysos}). Immediately to its west there lies molecular cloud No.~7 of \citet{yonekura2005}. It is accompanied by emission visible in the \textit{Herschel} bands and a peculiarly green point-like feature in the IRAC RGB image, that is, strong emission in the IRAC 4.5\,\ensuremath{\upmu\mathrm{m}} band. 

In the distribution of IRAC-identified cYSOs there is another notable cluster consisting of 12 candidates to the west of the H\,II region around $\upalpha_{2000} = 10$:34:27, $\updelta_{2000} = -58$:46:45 (orange circle in Fig.~\ref{fig:wise_ysos}). This region is devoid of mid-IR emission but is coincident with C\textsuperscript{18}O clump No.~1 of \citet{yonekura2005} and far-IR emission as traced by \textit{Herschel}. The cYSOs are aligned along the western ridge of the far-IR cloud as seen in the \textit{Herschel} image and even follow the shape of its filaments, broadly in the shape of an arrowhead pointing eastwards. The northern part is better aligned with the filament shape while the southern part is more randomly distributed around the filament itself. The border is also traced by the \citet{yonekura2005} \textsuperscript{12}CO intensity contours. With the WISE classification, however, there is nothing remarkable about that region. We find Class\,I candidate J103423.7--584531 to the north and Class\,II candidate J103424.6-584749 to the south, but no appearance of clustering.

North of the Gum\,31 bubble, around $\upalpha_{2000} = 10$:37:36, $\updelta_{2000} = -58$:26:36, there is a cluster of stars clearly discernible in the IRAC image (orange circle in Fig.~\ref{fig:wise_ysos}). It lies about 2\arcmin\ to the south-west of \citep{yonekura2005} C\textsuperscript{18}O clump No.~4. Around three dozen stars are seen within this group in projection and it is also associated with a number of \textit{Herschel}-identified point-like sources. Two of them, J103739.6--582756 and J103741.7--582629, are part of the sample analysed in Sect.~\ref{sec:herschelspitzer}. 
Few stars within or around this cluster are identified as cYSOs with WISE. J103736.3--582655 and J103741.9--582556 are the brightest stars in the IRAC images of the cluster and both identified as Class\,I candidates with WISE. There is one more Class\,I candidate and five Class\,II candidates distributed fairly evenly over the cluster.

There are several more minor clusters of $\sim 5$ cYSOs, notably always coincident with local maxima in the \citet{yonekura2005} C\textsuperscript{18}O maps.

\section{SED modelling for sources with both \textit{Herschel} and \textit{Spitzer} counterparts}
\label{sec:herschelspitzer}

Using our \textit{Herschel} and \textit{Spitzer} catalogues, we were able to identify those sources that are detected as point-like sources in both wavelength ranges. For \textit{Herschel} we applied the restriction that the sources had to be detected in at least three of the five bands, bringing the total number down from 91 sources detected in at least two bands to 59. This results in 16 sources overall that can be identified in at least three \textit{Herschel} bands and at least one of the IRAC bands. We then compared these identifications to the MIPS image and performed photometry as described in Sect.~\ref{sec:data-spitzer-mips} for those 6 sources where we could identify a MIPS counterpart. 

To extend the wavelength range of our observations, we additionally matched the point sources analysed here with sources from the 2MASS \citep{skrutskie2006} All-Sky Catalog of Point Sources \citep{cutri2003}. This was performed applying the same next-neighbour search as for the inter-band matching within the IRAC sources (Sect.~\ref{sec:data-spitzer-irac}), using only sources with quality flags A to D. We then repeated this process with the WISE catalogue, where we selected only those sources that had a signal-to-noise ratio larger than 5.
A detailed overview of all photometric data assembled is given in Table~\ref{tab:fluxes} in the online material. WISE 22\,\ensuremath{\upmu\mathrm{m}} and MIPS 24\,\ensuremath{\upmu\mathrm{m}} fluxes show some incongruity, however there is no underlying pattern as to what in the environs of the source may have influenced the photometry. This does not, however, influence the findings from the SED fits. In the two cases where the 22\,\ensuremath{\upmu\mathrm{m}} flux appears unduly high compared to the 24\,\ensuremath{\upmu\mathrm{m}} flux and an SED fit is performed, leaving out one or the other from the fit has little or no influence on the best-fit model.

\subsection{Modelling of the SEDs}
\label{sec:herschelspitzer-seds}

For SED-fitting we used the \href{http://caravan.astro.wisc.edu/protostars/sedfitter.php}{online tool} of \citet{robitaille2007}. This tool compares the input observational data with 200\,000 SED models for YSOs that were precomputed using a 2D radiative transfer code by \citet{whitney2003}. These models have a wide parameter space for the properties of the central object and its environment.

For the fits, the distance to all objects was assumed to be 2.3\,\ensuremath{\mathrm{kpc}} (cf.\ Sect.~\ref{sec:morphology}), and the interstellar extinction range was restricted to $A_V = 0 \dots 40$\,mag. We assumed an uncertainty of 20\% for all \textit{Herschel} fluxes. For 2MASS, \textit{Spitzer} and WISE fluxes in addition to the individual photometric measurement uncertainty as given in Table~\ref{tab:fluxes} we assumed a further systematic uncertainty of 10\% due e.\,g.\ to the reliability of flux calibration. For IRAC, photometry varies by up to 10\% due to the position of the point-like source within the detector array and though appropriate corrections were applied in the process, this is an additional source of uncertainty.

\begin{figure*}
 \centering
    \includegraphics[width=\textwidth]{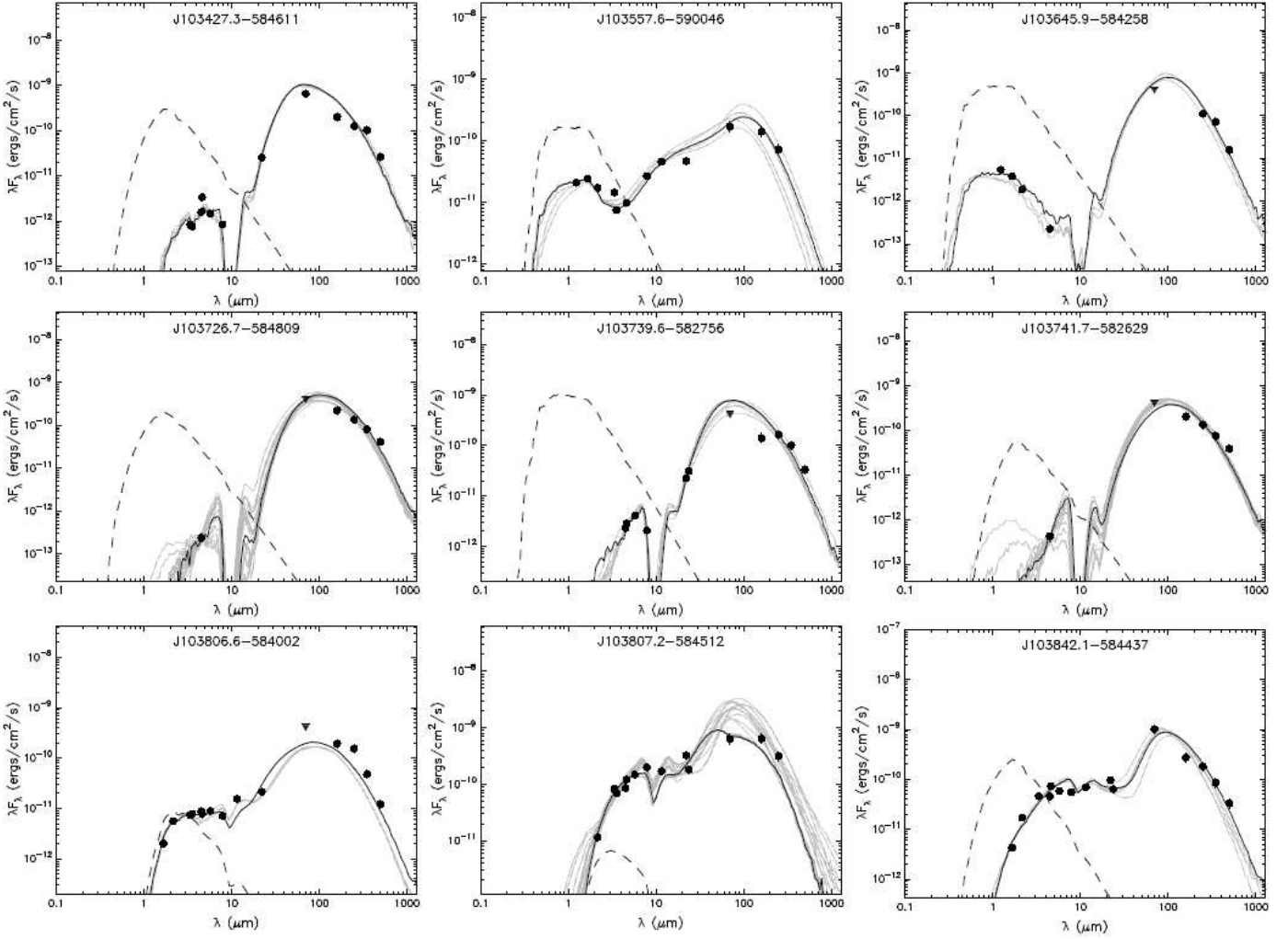}
 \caption{Spectral energy distributions of those objects for which we could determine fluxes with at least three \textit{Herschel} and one \textit{Spitzer} IRAC band. Filled circles mark the input fluxes. The black line shows the best fit, and the grey lines show subsequent good fits. The dashed line represents the stellar photosphere corresponding to the central source of the best fitting model, as it would appear in the absence of circumstellar dust (but including interstellar extinction).}
 \label{fig:seds}
\end{figure*}

In addition to the best-fit model, we show the range of possible parameters that can be derived from models within the range of $\chi^2\!/N - \chi_\mathrm{best}^2/N < 2$ (with $N$ representing the number of data points). These models are shown as grey lines in the plots in Fig.~\ref{fig:seds}. The resulting model parameters are listed in Table~\ref{tab:modell}. It gives the best-fit value together with the range constrained by the above $\chi^2$ criterion. The resulting SEDs are shown in Fig.~\ref{fig:seds}. We only use fits where $\chi^2\!/N$ for the best-fit model is smaller than or equal to 10.0.

\subsection{Results of SED modelling}

The results of SED fits can be highly ambiguous. Many of the stellar and circumstellar parameters are often poorly constrained because the models show a high degree of degeneracy \citep[e.\,g.][]{menshchikov1997}. We therefore restrict our analysis to a few selected parameters that can be relatively well determined from these fits. These are the total luminosity, the stellar mass, and the mass of the circumstellar disk and envelope.

The best-fit masses as listed in Table~\ref{tab:modell} are between 1.7\,\ensuremath{M_\odot} and 6.6\,\ensuremath{M_\odot} and even the extremes of the ranges do not exceed 1.2\,\ensuremath{M_\odot} to 7.1\,\ensuremath{M_\odot}. The majority of luminosities are to be found in a range of $\sim 100$ -- 300\,\ensuremath{L_\odot}, with two notable exceptions well below that at 38\,\ensuremath{L_\odot} and 42\,\ensuremath{L_\odot}, respectively, and one exceptionally luminous source at 890\,\ensuremath{L_\odot} best-fit value, corresponding with the highest best-fit stellar mass in our sample. 
Whereas 4 of the 10 sources sampled here exhibit best-fit envelope masses of 190\,\ensuremath{M_\odot} or higher, 3 are at $\leq 100$\,\ensuremath{M_\odot} and two lower than 50\,\ensuremath{M_\odot}. The disk masses span a range of about one order of magnitude between $\sim 0.01\,\ensuremath{M_\odot}$ and $\sim 0.1\,\ensuremath{M_\odot}$. The highest-mass star in the sample is the notable exeption with a disk mass of $\sim 0.001\,\ensuremath{M_\odot}$.

In a large-scale view it is immediately noticeable that all but two of the sources for which \textit{Herschel} counterparts to \textit{Spitzer} point-like sources are detected are to be found within the Gum\,31 bubble. Although the field of our study stretches further out especially to the west, only two sources are found outside the bubble. These are J103557.6--590046 and J103427.3--584611.

J103806.6--584002 is remarkable in that contrary to the vast majority of objects it is not located in the rim of the bubble but within the bubble itself, being the only specimen in our sample. It is placed in the very head of a pillar-like filament that extends from the northern rim of the bubble into it (cf.\ Fig.~\ref{fig:spitzer-herschel_266}). With the methods employed in Sect.~\ref{sec:ysos} we classify it as a WISE Class\,I candidate and an IRAC cYSO. In the \textit{Herschel} images the filament is rather faint, but J103806.6--584002 itself is clearly visible as a point-like source.

\section{Sources of HH jets}
\label{sec:hhjets}

In an earlier paper \citep{ohlendorf2012} we have traced a number of Herbig-Haro jets identified in the Carina Nebula complex by \citet{smith2010_jets} back to their protostellar sources. \citet{smith2010_jets} also identify two HH jets and two HH jet candidates in the Gum\,31 bubble. It should be remarked that the HST images on which the \citet{smith2010_jets} study is based cover only a very small area of the entire Gum\,31 region and the following analysis only represents a very small section of the area covered in the rest of the paper. Due to the small sample size and its very limited spatial dimensions, we cannot draw any conclusions about the distribution or likely number of all jet-driving protostars within the Gum\,31 region.

\begin{figure*}
 \centering
 \includegraphics[width=\textwidth]{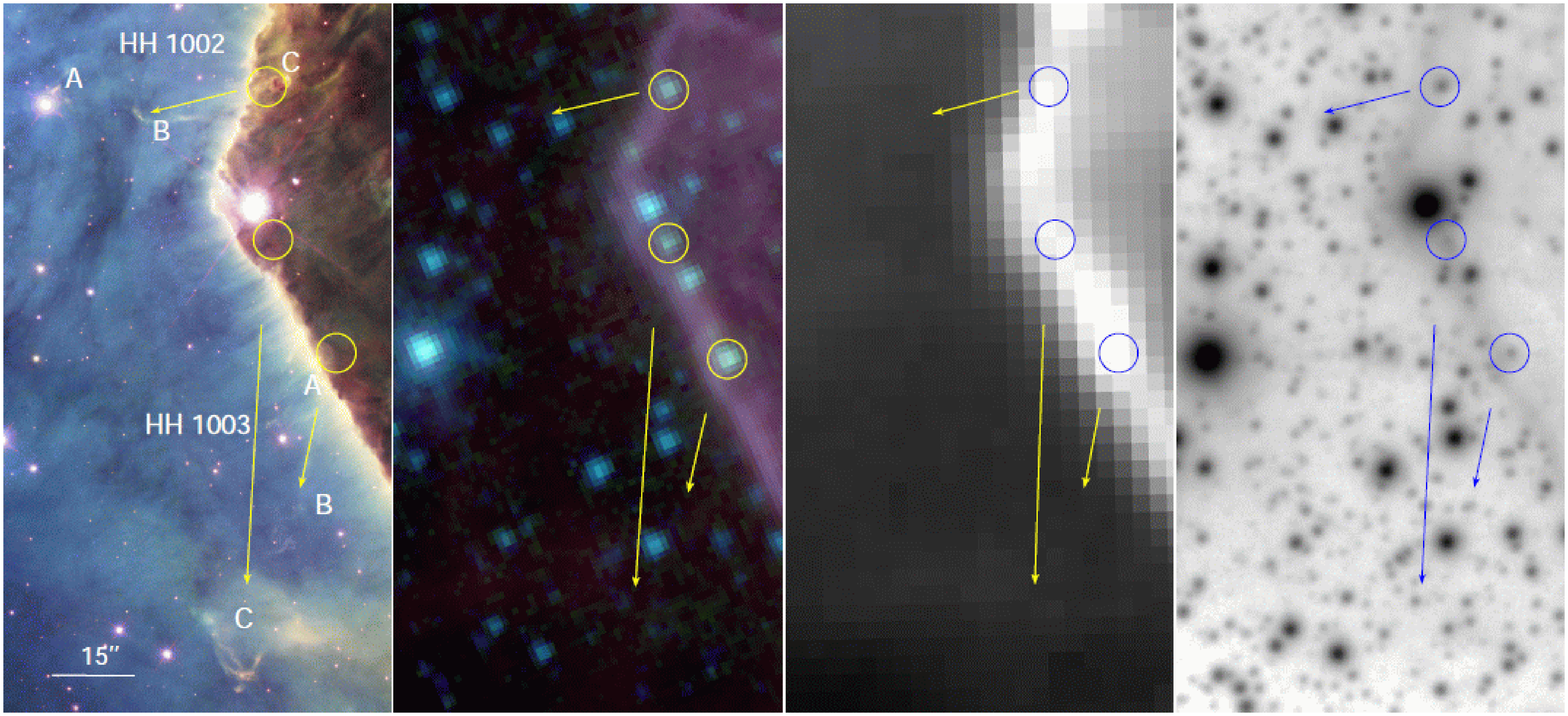}
 \caption{The Herbig-Haro jets and their probable sources as seen in four different wavelengths. From left to right: Hubble Space Telescope RGB image (red: WFPC2 S\,II filter (673\,nm), green: ACS H$\upalpha$ +N\,II filter (658\,nm), blue: WFPC2 O\,III filter (502\,nm); image credit: NASA/ESA/Hubble Heritage Team (STScI/AURA)), \textit{Spitzer} IRAC RGB image with 3.6\,\ensuremath{\upmu\mathrm{m}} in blue, 4.5\,\ensuremath{\upmu\mathrm{m}} in green and 8.0\,\ensuremath{\upmu\mathrm{m}} in red, \textit{Herschel} PACS 170\,\ensuremath{\upmu\mathrm{m}} image and VISTA \emph{H}-band image. The yellow or blue arrows indicate the broad shape and direction of the outflow in question, white letters mark their features as described by \citet{smith2010_jets}. The probable IR-identified sources are marked at their \textit{Spitzer} coordinates with circles in yellow or blue.}
 \label{fig:images_hhs}
\end{figure*}

As can be seen in Fig.~\ref{fig:images_hhs}, the jet HH\,1002 is almost perpendicular to the ionisation front. It shows a number of distinct features, marked by \citet{smith2010_jets} as A, B and C and labelled thus in our figure. \citet{smith2010_jets} remark that in the 2MASS images they detect a very likely source located at $\upalpha_{2000} = 10$:36:53.9, $\updelta_{2000} = -58$:37:19. This matches almost exactly with the location of the object identified in IRAC images that is most likely to be the source of the jet: J103654.0--583720, which we classified as being a cYSO through its WISE and IRAC infrared excesses in Sect.~\ref{sec:ysos-identification}. The \textit{Herschel} observations trace the ionisation front well, but show no point-like source to be coincident with J103654.0--583720. Therefore it could not be included in the SED-based analysis in Sect.~\ref{sec:herschelspitzer}.

HH\,1003 has a more complicated structure and \citet{smith2010_jets} discuss it as possibly being a two-part object, made out of two jets in close association. Tracing back the direction of the bow shocks in Fig.~\ref{fig:images_hhs}, we find two point-like sources in the IRAC images that are very probably the sources of two different jets. One of them, J103652.4--583809, associated with features A and B, is identified as a cYSO from the WISE and IRAC colour-colour diagrams, while the other is an IRAC cYSO. Both are coincident with faint sources in the 2MASS images. Again, the \textit{Herschel} image shows no point-like sources.

The two candidate outflows, HH\,c-1 and HH\,c-2 (not to be confused with HH\,c-1 and HH\,c-2 within the extent of the central Carina Nebula which were included in the study of \citeauthor{ohlendorf2012}\ \citeyear{ohlendorf2012}), could not be traced back to any IR sources within the scope of our study. Following the axes of the jets, we cannot identify any likely emitting sources within their immediate surroundings.

\section{Discussion and Conclusions}
\label{sec:conclusions}

In this paper we analysed \textit{Spitzer}, WISE, and \textit{Herschel} data to investigate the cloud structure and the young stellar population in and around the Gum\,31 nebula. These data provide considerably better sensitivity and spatial resolution than the previously available data sets.

The \textit{Herschel} far-IR maps show that the bubble surrounding the Gum\,31 nebula is connected to the central parts of the Carina Nebula. This adds strong direct support to the assumption that Gum\,31 is actually part of the Carina Nebula complex, as is also suggested by the matching C\textsuperscript{18}O radial velocities measured by \citet{yonekura2005} and other recent distance determinations (cf.\ Sect.~\ref{sec:intro}).

The bubble itself has a very sharp western edge, where the column density derived from \textit{Herschel} measurements rises abruptly by at least an order of magnitude. The dust temperatures range from $\la 20$\,K in the dark clouds surrounding the H\,II region to $\approx 30$\,K in the H\,II region and up to  $\approx 40$\,K near the location of the O-type stars.

The very small excess fraction seen for the mid-IR sources in the central stellar cluster NGC\,3324 ($\approx$ 0.5\%) suggests that it is at least several Myr old already.
For the whole Gum\,31 region, the \textit{Spitzer} and WISE data reveal about 300/660 cYSOs. These objects are most likely Class\,I protostars or Class\,II sources of solar to intermediate mass.
The 59 far-IR point-like sources revealed by the \textit{Herschel} data are either pre-stellar cores or embedded (Class~0) protostars, i.\,e.\ trace a younger population of currently forming stars. The spatial distribution of the cYSOs is highly non-uniform. As we expect a contamination of around 18\% most of the widely distributed YSO population we see is probably due to back- and foreground stars as those would be expected to be distributed homogeneously. Many cYSOs are found in rather compact clusterings, and a considerable number is found at the inner edge of the dusty bubble surrounding the H\,II region. The \textit{Herschel}-identified point-like sources in particular trace the edge of the bubble.

This led us to assume a `collect and collapse' scenario driven by the O stars within the cluster \citep{whitworth1994}, resulting in an expected bubble size of 9 -- 11\,\ensuremath{\mathrm{pc}} for an age of 1.5 -- 2.0\,\ensuremath{\mathrm{Myr}}. This agrees very well with the $\sim 10$\,\ensuremath{\mathrm{pc}} diameter observed. We find four cYSOs in the very tips of small pillars extending from the bubble rim (cf.\ Fig.~\ref{fig:spitzer-herschel_266}), suggesting radiative triggering processes very similar to what is observed in the South Pillars \citep{smith2010_spitzer, gaczkowski2012}.

We conclude that two different modes of triggered star-formation occur simultaneously in the Gum\,31 region: `collect and collapse', as evidenced by the bubble size and the cYSOs in its rim, and radiative triggering, as evidenced by cYSOs in the heads of pillars.

We construct the near- to far-infrared SEDs of 17 cYSOs and estimate basic stellar and circumstellar parameters by comparison to radiative-transfer models with good-quality fits for 10 of them. All these cYSO s are of moderate luminosity ($L \la 900\,\ensuremath{L_\odot}$), clearly suggesting that they are low- or intermediate-mass objects ($M \la 7\,\ensuremath{M_\odot}$). This agrees with the results from our analysis of the cYSOs in the central parts of the Carina Nebula complex \citep{gaczkowski2012}, where we found that no high-mass stars are currently forming.

We identify the driving sources of two Herbig-Haro jets in the western rim of the Gum\,31 bubble. These sources are also identified as cYSOs by applying colour-selection criteria to IRAC or WISE photometry data.

From the total number of cYSOs observed and the IMF \citep{kroupa2002} we can estimate a total young stellar population for the Gum\,31 region. Our detection limit for cYSOs is about 1\,\ensuremath{M_\odot} and following the IMF there should be eight times as many stars below this mass ($> 0.1$\,\ensuremath{M_\odot}) as above it. Correcting the number of cYSOs given in Sect.~\ref{sec:ysos-identification-wise} for the contamination estimated there, this gives a number of $\sim 5000$ young stellar objects in the region.

A more detailed investigation of the star formation history in this area requires a reliable identification of the individual young stars. While the infrared data presented in this paper can reveal protostars and young stars with circumstellar disks, most of the slightly older ($\gtrsim 2$\,Myr old) stars cannot be detected by infrared excesses.
Our very recent \textit{Chandra} X-ray observations of the Gum\,31 region, near-IR observations with VISTA and other ongoing observations will allow us to finally identify these stars, too. This will constitute the basis for a comprehensive multi-wavelength study of this interesting region, in a way similar to the recent studies of the young stellar populations in the central parts of the Carina Nebula \citep[see][]{townsley2011, preibisch2011_cccp, wang2011, wolk2011, feigelson2011}.

\begin{acknowledgements}

This work was supported by the German \emph{Deut\-sche For\-schungs\-ge\-mein\-schaft, DFG\/} project number 569/9-1. Additional support came from funds from the Munich Cluster of Excellence ``Origin and Structure of the Universe''. 

The authors would like to thank the \textit{Spitzer} Science Center Helpdesk for large amounts of support while working with MOPEX.

This work is based in part on archival data obtained with the \textit{Spitzer} Space Telescope, which is operated by the Jet Propulsion Laboratory, California Institute of Technology under a contract with NASA.

This publication makes use of data obtained with the \textit{Herschel} spacecraft. The \textit{Herschel} spacecraft was designed, built, tested, and launched under a contract to ESA managed by the \textit{Herschel}/\textit{Planck} Project team by an industrial consortium under the overall responsibility of the prime contractor Thales Alenia Space (Cannes), and including Astrium (Friedrichshafen) responsible for the payload module and for system testing at spacecraft level, Thales Alenia Space (Turin) responsible for the service module, and Astrium (Toulouse) responsible for the telescope, with in excess of a hundred subcontractors.

This publication makes use of data products from the Two Micron All Sky Survey, which is a joint project of the University of Massachusetts and the Infrared Processing and Analysis Center/California Institute of Technology, funded by the National Aeronautics and Space Administration and the National Science Foundation.

This publication makes use of data products from the Wide-field Infrared Survey Explorer, which is a joint project of the University of California, Los Angeles, and the Jet Propulsion Laboratory/California Institute of Technology, funded by the National Aeronautics and Space Administration.

The science data reduction for VISTA up to the creation of the final tile was performed by the \href{http://casu.ast.cam.ac.uk/surveys-projects/vista/data-processing}{Cambridge Astronomy Survey Unit}.

The Digitized Sky Survey was produced at the Space Telescope Science Institute under U.S. Government grant NAG W-2166. The images of these surveys are based on photographic data obtained using the Oschin Schmidt Telescope on Palomar Mountain and the UK Schmidt Telescope. The plates were processed into the present compressed digital form with the permission of these institutions. 

This research has made use of NASA's Astrophysics Data System Bibliographic Services.

This research has made use of the SIMBAD database and the VizieR catalogue access tool, operated at CDS, Strasbourg, France.

\end{acknowledgements}

\bibliographystyle{aa}
\bibliography{aa201220218.bib}

\begin{thebibliography}{68}
\expandafter\ifx\csname natexlab\endcsname\relax\def\natexlab#1{#1}\fi

\bibitem[{{Allen} {et~al.}(2004){Allen}, {Calvet}, {D'Alessio}, {Merin},
  {Hartmann}, {Megeath}, {Gutermuth}, {Muzerolle}, {Pipher}, {Myers}, \&
  {Fazio}}]{allen2004}
{Allen}, L.~E., {Calvet}, N., {D'Alessio}, P., {et~al.} 2004, \apjs, 154, 363

\bibitem[{{Baraffe} {et~al.}(1998){Baraffe}, {Chabrier}, {Allard}, \&
  {Hauschildt}}]{baraffe1998}
{Baraffe}, I., {Chabrier}, G., {Allard}, F., \& {Hauschildt}, P.~H. 1998, \aap,
  337, 403

\bibitem[{{Barnes} {et~al.}(2010){Barnes}, {Yonekura}, {Ryder}, {Hopkins},
  {Miyamoto}, {Furukawa}, \& {Fukui}}]{barnes2010}
{Barnes}, P.~J., {Yonekura}, Y., {Ryder}, S.~D., {et~al.} 2010, \mnras, 402, 73

\bibitem[{{Baumgardt} {et~al.}(2000){Baumgardt}, {Dettbarn}, \&
  {Wielen}}]{baumgardt2000}
{Baumgardt}, H., {Dettbarn}, C., \& {Wielen}, R. 2000, \aaps, 146, 251

\bibitem[{{Cappa} {et~al.}(2008){Cappa}, {Niemela}, {Amor{\'{\i}}n}, \&
  {Vasquez}}]{cappa2008}
{Cappa}, C., {Niemela}, V.~S., {Amor{\'{\i}}n}, R., \& {Vasquez}, J. 2008,
  \aap, 477, 173

\bibitem[{{Carraro} {et~al.}(2001){Carraro}, {Patat}, \&
  {Baumgardt}}]{carraro2001}
{Carraro}, G., {Patat}, F., \& {Baumgardt}, H. 2001, \aap, 371, 107

\bibitem[{{Clari{\'a}}(1977)}]{claria1977}
{Clari{\'a}}, J.~J. 1977, \aaps, 27, 145

\bibitem[{{Cutri} \& {al.}(2012)}]{cutri2012}
{Cutri}, R.~M. \& {al.} 2012, VizieR Online Data Catalog, 2311

\bibitem[{{Cutri} {et~al.}(2003){Cutri}, {Skrutskie}, {van Dyk}, {Beichman},
  {Carpenter}, {Chester}, {Cambresy}, {Evans}, {Fowler}, {Gizis}, {Howard},
  {Huchra}, {Jarrett}, {Kopan}, {Kirkpatrick}, {Light}, {Marsh}, {McCallon},
  {Schneider}, {Stiening}, {Sykes}, {Weinberg}, {Wheaton}, {Wheelock}, \&
  {Zacarias}}]{cutri2003}
{Cutri}, R.~M., {Skrutskie}, M.~F., {van Dyk}, S., {et~al.} 2003, VizieR Online
  Data Catalog, 2246

\bibitem[{{Cutri} {et~al.}(2012){Cutri}, {Wright}, {Conrow}, {Bauer},
  {Benford}, {Brandenburg}, {Dailey}, {Eisenhardt}, {Evans}, {Fajardo-Acosta},
  {Fowler}, {Gelino}, {Grillmair}, {Harbut}, {Hoffman}, {Jarrett},
  {Kirkpatrick}, {Leisawitz}, {Liu}, {Mainzer}, {Marsh}, {Masci}, {McCallon},
  {Padgett}, {Ressler}, {Royer}, {Skrutskie}, {Stanford}, {Wyatt}, {Tholen},
  {Tsai}, {Wachter}, {Wheelock}, {Yan}, {Alles}, {Beck}, {Grav}, {Masiero},
  {McCollum}, {McGehee}, {Papin}, \&
  {Wittman}}]{cutriEtal2012_wiseExplanatorySupplement}
{Cutri}, R.~M., {Wright}, E.~L., {Conrow}, T., {et~al.} 2012, {Explanatory
  Supplement to the WISE All-Sky Data Release Products}, Tech. rep.

\bibitem[{{Dalton} {et~al.}(2006){Dalton}, {Caldwell}, {Ward}, {Whalley},
  {Woodhouse}, {Edeson}, {Clark}, {Beard}, {Gallie}, {Todd}, {Strachan},
  {Bezawada}, {Sutherland}, \& {Emerson}}]{dalton2006}
{Dalton}, G.~B., {Caldwell}, M., {Ward}, A.~K., {et~al.} 2006, in Society of
  Photo-Optical Instrumentation Engineers (SPIE) Conference Series, Vol. 6269

\bibitem[{{Dewangan} {et~al.}(2012){Dewangan}, {Ojha}, {Anandarao}, {Ghosh}, \&
  {Chakraborti}}]{dewangan2012}
{Dewangan}, L.~K., {Ojha}, D.~K., {Anandarao}, B.~G., {Ghosh}, S.~K., \&
  {Chakraborti}, S. 2012, arXiv:1207.6842v1

\bibitem[{{Diolaiti} {et~al.}(2000){Diolaiti}, {Bendinelli}, {Bonaccini},
  {Close}, {Currie}, \& {Parmeggiani}}]{diolaiti2000}
{Diolaiti}, E., {Bendinelli}, O., {Bonaccini}, D., {et~al.} 2000, \aaps, 147,
  335

\bibitem[{{Dullemond} \& {Dominik}(2004)}]{dullemond2004}
{Dullemond}, C.~P. \& {Dominik}, C. 2004, \aap, 417, 159

\bibitem[{{Dutra} {et~al.}(2003){Dutra}, {Bica}, {Soares}, \&
  {Barbuy}}]{dutra2003}
{Dutra}, C.~M., {Bica}, E., {Soares}, J., \& {Barbuy}, B. 2003, \aap, 400, 533

\bibitem[{{Emerson} \& {Sutherland}(2010)}]{emerson2010}
{Emerson}, J. \& {Sutherland}, W. 2010, The Messenger, 139, 2

\bibitem[{{Fang} {et~al.}(2012){Fang}, {van Boekel}, {King}, {Henning},
  {Bouwman}, {Doi}, {Okamoto}, {Roccatagliata}, \&
  {Sicilia-Aguilar}}]{fang2012}
{Fang}, M., {van Boekel}, R., {King}, R.~R., {et~al.} 2012, \aap, 539, A119

\bibitem[{{Fa{\'u}ndez} {et~al.}(2004){Fa{\'u}ndez}, {Bronfman}, {Garay},
  {Chini}, {Nyman}, \& {May}}]{faundez2004}
{Fa{\'u}ndez}, S., {Bronfman}, L., {Garay}, G., {et~al.} 2004, \aap, 426, 97

\bibitem[{{Fazio} {et~al.}(2004){Fazio}, {Hora}, {Allen}, {Ashby}, {Barmby},
  {Deutsch}, {Huang}, {Kleiner}, {Marengo}, {Megeath}, {Melnick}, {Pahre},
  {Patten}, {Polizotti}, {Smith}, {Taylor}, {Wang}, {Willner}, {Hoffmann},
  {Pipher}, {Forrest}, {McMurty}, {McCreight}, {McKelvey}, {McMurray}, {Koch},
  {Moseley}, {Arendt}, {Mentzell}, {Marx}, {Losch}, {Mayman}, {Eichhorn},
  {Krebs}, {Jhabvala}, {Gezari}, {Fixsen}, {Flores}, {Shakoorzadeh}, {Jungo},
  {Hakun}, {Workman}, {Karpati}, {Kichak}, {Whitley}, {Mann}, {Tollestrup},
  {Eisenhardt}, {Stern}, {Gorjian}, {Bhattacharya}, {Carey}, {Nelson},
  {Glaccum}, {Lacy}, {Lowrance}, {Laine}, {Reach}, {Stauffer}, {Surace},
  {Wilson}, {Wright}, {Hoffman}, {Domingo}, \& {Cohen}}]{fazio2004}
{Fazio}, G.~G., {Hora}, J.~L., {Allen}, L.~E., {et~al.} 2004, \apjs, 154, 10

\bibitem[{{Feigelson} {et~al.}(2011){Feigelson}, {Getman}, {Townsley}, {Broos},
  {Povich}, {Garmire}, {King}, {Montmerle}, {Preibisch}, {Smith}, {Stassun},
  {Wang}, {Wolk}, \& {Zinnecker}}]{feigelson2011}
{Feigelson}, E.~D., {Getman}, K.~V., {Townsley}, L.~K., {et~al.} 2011, \apjs,
  194, 9

\bibitem[{{Flaherty} {et~al.}(2007){Flaherty}, {Pipher}, {Megeath}, {Winston},
  {Gutermuth}, {Muzerolle}, {Allen}, \& {Fazio}}]{flaherty2007}
{Flaherty}, K.~M., {Pipher}, J.~L., {Megeath}, S.~T., {et~al.} 2007, \apj, 663,
  1069

\bibitem[{{Forte}(1976)}]{forte1976}
{Forte}, J.~C. 1976, \aaps, 25, 271

\bibitem[{{Gaczkowski} {et~al.}(2013){Gaczkowski}, {Preibisch}, {Ratzka},
  {Roccatagliata}, {Ohlendorf}, \& {Zinnecker}}]{gaczkowski2012}
{Gaczkowski}, B., {Preibisch}, T., {Ratzka}, T., {et~al.} 2013, \aap, 549, A67

\bibitem[{{Griffin} {et~al.}(2010){Griffin}, {Abergel}, {Abreu}, {Ade},
  {Andr{\'e}}, {Augueres}, {Babbedge}, {Bae}, {Baillie}, {Baluteau}, {Barlow},
  {Bendo}, {Benielli}, {Bock}, {Bonhomme}, {Brisbin}, {Brockley-Blatt},
  {Caldwell}, {Cara}, {Castro-Rodriguez}, {Cerulli}, {Chanial}, {Chen},
  {Clark}, {Clements}, {Clerc}, {Coker}, {Communal}, {Conversi}, {Cox},
  {Crumb}, {Cunningham}, {Daly}, {Davis}, {de Antoni}, {Delderfield}, {Devin},
  {di Giorgio}, {Didschuns}, {Dohlen}, {Donati}, {Dowell}, {Dowell}, {Duband},
  {Dumaye}, {Emery}, {Ferlet}, {Ferrand}, {Fontignie}, {Fox}, {Franceschini},
  {Frerking}, {Fulton}, {Garcia}, {Gastaud}, {Gear}, {Glenn}, {Goizel},
  {Griffin}, {Grundy}, {Guest}, {Guillemet}, {Hargrave}, {Harwit}, {Hastings},
  {Hatziminaoglou}, {Herman}, {Hinde}, {Hristov}, {Huang}, {Imhof}, {Isaak},
  {Israelsson}, {Ivison}, {Jennings}, {Kiernan}, {King}, {Lange}, {Latter},
  {Laurent}, {Laurent}, {Leeks}, {Lellouch}, {Levenson}, {Li}, {Li},
  {Lilienthal}, {Lim}, {Liu}, {Lu}, {Madden}, {Mainetti}, {Marliani}, {McKay},
  {Mercier}, {Molinari}, {Morris}, {Moseley}, {Mulder}, {Mur}, {Naylor},
  {Nguyen}, {O'Halloran}, {Oliver}, {Olofsson}, {Olofsson}, {Orfei}, {Page},
  {Pain}, {Panuzzo}, {Papageorgiou}, {Parks}, {Parr-Burman}, {Pearce},
  {Pearson}, {P{\'e}rez-Fournon}, {Pinsard}, {Pisano}, {Podosek}, {Pohlen},
  {Polehampton}, {Pouliquen}, {Rigopoulou}, {Rizzo}, {Roseboom}, {Roussel},
  {Rowan-Robinson}, {Rownd}, {Saraceno}, {Sauvage}, {Savage}, {Savini},
  {Sawyer}, {Scharmberg}, {Schmitt}, {Schneider}, {Schulz}, {Schwartz},
  {Shafer}, {Shupe}, {Sibthorpe}, {Sidher}, {Smith}, {Smith}, {Smith},
  {Spencer}, {Stobie}, {Sudiwala}, {Sukhatme}, {Surace}, {Stevens}, {Swinyard},
  {Trichas}, {Tourette}, {Triou}, {Tseng}, {Tucker}, {Turner}, {Vaccari},
  {Valtchanov}, {Vigroux}, {Virique}, {Voellmer}, {Walker}, {Ward}, {Waskett},
  {Weilert}, {Wesson}, {White}, {Whitehouse}, {Wilson}, {Winter}, {Woodcraft},
  {Wright}, {Xu}, {Zavagno}, {Zemcov}, {Zhang}, \& {Zonca}}]{griffin2010}
{Griffin}, M.~J., {Abergel}, A., {Abreu}, A., {et~al.} 2010, \aap, 518, L3

\bibitem[{{Hoffleit}(1953)}]{hoffleit1953}
{Hoffleit}, D. 1953, Annals of Harvard College Observatory, 119, 37

\bibitem[{{Irwin} {et~al.}(2004){Irwin}, {Lewis}, {Hodgkin}, {Bunclark},
  {Evans}, {McMahon}, {Emerson}, {Stewart}, \& {Beard}}]{irwin2004}
{Irwin}, M.~J., {Lewis}, J., {Hodgkin}, S., {et~al.} 2004, in Society of
  Photo-Optical Instrumentation Engineers (SPIE) Conference Series, ed.
  {P.~J.~Quinn \& A.~Bridger}, Vol. 5493, 411--422

\bibitem[{{Kharchenko} {et~al.}(2005){Kharchenko}, {Piskunov}, {R{\"o}ser},
  {Schilbach}, \& {Scholz}}]{kharchenko2005}
{Kharchenko}, N.~V., {Piskunov}, A.~E., {R{\"o}ser}, S., {Schilbach}, E., \&
  {Scholz}, R.-D. 2005, \aap, 438, 1163

\bibitem[{{Koenig} {et~al.}(2012){Koenig}, {Leisawitz}, {Benford}, {Rebull},
  {Padgett}, \& {Assef}}]{koenig2012}
{Koenig}, X.~P., {Leisawitz}, D.~T., {Benford}, D.~J., {et~al.} 2012, \apj,
  744, 130

\bibitem[{{Kramer} {et~al.}(2008){Kramer}, {Cubick}, {R{\"o}llig}, {Sun},
  {Yonekura}, {Aravena}, {Bensch}, {Bertoldi}, {Bronfman}, {Fujishita},
  {Fukui}, {Graf}, {Hitschfeld}, {Honingh}, {Ito}, {Jakob}, {Jacobs}, {Klein},
  {Koo}, {May}, {Miller}, {Miyamoto}, {Mizuno}, {Onishi}, {Park}, {Pineda},
  {Rabanus}, {Sasago}, {Schieder}, {Simon}, {Stutzki}, {Volgenau}, \&
  {Yamamoto}}]{kramer2008}
{Kramer}, C., {Cubick}, M., {R{\"o}llig}, M., {et~al.} 2008, \aap, 477, 547

\bibitem[{{Kroupa}(2002)}]{kroupa2002}
{Kroupa}, P. 2002, Science, 295, 82

\bibitem[{{Kudritzki} {et~al.}(1999){Kudritzki}, {Puls}, {Lennon}, {Venn},
  {Reetz}, {Najarro}, {McCarthy}, \& {Herrero}}]{kudritzki1999}
{Kudritzki}, R.~P., {Puls}, J., {Lennon}, D.~J., {et~al.} 1999, \aap, 350, 970

\bibitem[{{Lada}(1987)}]{lada1987}
{Lada}, C.~J. 1987, in IAU Symposium, Vol. 115, Star Forming Regions, ed.
  {M.~Peimbert \& J.~Jugaku}, 1--17

\bibitem[{{Lyng{\aa}} \& {Hansson}(1972)}]{lyngaa1972}
{Lyng{\aa}}, G. \& {Hansson}, N. 1972, \aaps, 6, 327

\bibitem[{{Ma{\'{\i}}z-Apell{\'a}niz}
  {et~al.}(2004){Ma{\'{\i}}z-Apell{\'a}niz}, {Walborn}, {Galu{\'e}}, \&
  {Wei}}]{maizApellaniz2004}
{Ma{\'{\i}}z-Apell{\'a}niz}, J., {Walborn}, N.~R., {Galu{\'e}}, H.~{\'A}., \&
  {Wei}, L.~H. 2004, \apjs, 151, 103

\bibitem[{{Makovoz} \& {Marleau}(2005)}]{makovoz2005}
{Makovoz}, D. \& {Marleau}, F.~R. 2005, \pasp, 117, 1113

\bibitem[{{Men'shchikov} \& {Henning}(1997)}]{menshchikov1997}
{Men'shchikov}, A.~B. \& {Henning}, T. 1997, \aap, 318, 879

\bibitem[{{Molinari} {et~al.}(2011){Molinari}, {Schisano}, {Faustini},
  {Pestalozzi}, {di Giorgio}, \& {Liu}}]{molinari2011}
{Molinari}, S., {Schisano}, E., {Faustini}, F., {et~al.} 2011, \aap, 530, A133

\bibitem[{{Mottram} {et~al.}(2007){Mottram}, {Hoare}, {Lumsden}, {Oudmaijer},
  {Urquhart}, {Sheret}, {Clarke}, \& {Allsopp}}]{mottram2007}
{Mottram}, J.~C., {Hoare}, M.~G., {Lumsden}, S.~L., {et~al.} 2007, \aap, 476,
  1019

\bibitem[{{Nesterov} {et~al.}(1995){Nesterov}, {Kuzmin}, {Ashimbaeva},
  {Volchkov}, {R{\"o}ser}, \& {Bastian}}]{nesterov1995}
{Nesterov}, V.~V., {Kuzmin}, A.~V., {Ashimbaeva}, N.~T., {et~al.} 1995, \aaps,
  110, 367

\bibitem[{{Ohlendorf} {et~al.}(2012){Ohlendorf}, {Preibisch}, {Gaczkowski},
  {Ratzka}, {Grellmann}, \& {McLeod}}]{ohlendorf2012}
{Ohlendorf}, H., {Preibisch}, T., {Gaczkowski}, B., {et~al.} 2012, \aap, 540,
  A81

\bibitem[{{Oliveira} {et~al.}(2009){Oliveira}, {Mer{\'{\i}}n}, {Pontoppidan},
  {van Dishoeck}, {Overzier}, {Hern{\'a}ndez}, {Sicilia-Aguilar}, {Eiroa}, \&
  {Montesinos}}]{oliveira2009}
{Oliveira}, I., {Mer{\'{\i}}n}, B., {Pontoppidan}, K.~M., {et~al.} 2009, \apj,
  691, 672

\bibitem[{{Pickles} \& {Depagne}(2010)}]{pickles2010}
{Pickles}, A. \& {Depagne}, {\'E}. 2010, \pasp, 122, 1437

\bibitem[{{Poglitsch} {et~al.}(2010){Poglitsch}, {Waelkens}, {Geis},
  {Feuchtgruber}, {Vandenbussche}, {Rodriguez}, {Krause}, {Renotte}, {van
  Hoof}, {Saraceno}, {Cepa}, {Kerschbaum}, {Agn{\`e}se}, {Ali}, {Altieri},
  {Andreani}, {Augueres}, {Balog}, {Barl}, {Bauer}, {Belbachir}, {Benedettini},
  {Billot}, {Boulade}, {Bischof}, {Blommaert}, {Callut}, {Cara}, {Cerulli},
  {Cesarsky}, {Contursi}, {Creten}, {De Meester}, {Doublier}, {Doumayrou},
  {Duband}, {Exter}, {Genzel}, {Gillis}, {Gr{\"o}zinger}, {Henning},
  {Herreros}, {Huygen}, {Inguscio}, {Jakob}, {Jamar}, {Jean}, {de Jong},
  {Katterloher}, {Kiss}, {Klaas}, {Lemke}, {Lutz}, {Madden}, {Marquet},
  {Martignac}, {Mazy}, {Merken}, {Montfort}, {Morbidelli}, {M{\"u}ller},
  {Nielbock}, {Okumura}, {Orfei}, {Ottensamer}, {Pezzuto}, {Popesso},
  {Putzeys}, {Regibo}, {Reveret}, {Royer}, {Sauvage}, {Schreiber}, {Stegmaier},
  {Schmitt}, {Schubert}, {Sturm}, {Thiel}, {Tofani}, {Vavrek}, {Wetzstein},
  {Wieprecht}, \& {Wiezorrek}}]{poglitsch2010}
{Poglitsch}, A., {Waelkens}, C., {Geis}, N., {et~al.} 2010, \aap, 518, L2

\bibitem[{{Preibisch} {et~al.}(2011{\natexlab{a}}){Preibisch}, {Hodgkin},
  {Irwin}, {Lewis}, {King}, {McCaughrean}, {Zinnecker}, {Townsley}, \&
  {Broos}}]{preibisch2011_cccp}
{Preibisch}, T., {Hodgkin}, S., {Irwin}, M., {et~al.} 2011{\natexlab{a}},
  \apjs, 194, 10

\bibitem[{{Preibisch} {et~al.}(2011{\natexlab{b}}){Preibisch}, {Ratzka},
  {Kuderna}, {Ohlendorf}, {King}, {Hodgkin}, {Irwin}, {Lewis}, {McCaughrean},
  \& {Zinnecker}}]{preibisch2011_hawki}
{Preibisch}, T., {Ratzka}, T., {Kuderna}, B., {et~al.} 2011{\natexlab{b}},
  \aap, 530, A34

\bibitem[{{Preibisch} {et~al.}(2012){Preibisch}, {Roccatagliata}, {Gaczkowski},
  \& {Ratzka}}]{preibisch2012}
{Preibisch}, T., {Roccatagliata}, V., {Gaczkowski}, B., \& {Ratzka}, T. 2012,
  \aap, 541, A132

\bibitem[{{Preibisch} {et~al.}(2011{\natexlab{c}}){Preibisch}, {Schuller},
  {Ohlendorf}, {Pekruhl}, {Menten}, \& {Zinnecker}}]{preibisch2011_laboca}
{Preibisch}, T., {Schuller}, F., {Ohlendorf}, H., {et~al.} 2011{\natexlab{c}},
  \aap, 525, A92

\bibitem[{{Przybilla} {et~al.}(2006){Przybilla}, {Butler}, {Becker}, \&
  {Kudritzki}}]{przybilla2006}
{Przybilla}, N., {Butler}, K., {Becker}, S.~R., \& {Kudritzki}, R.~P. 2006,
  \aap, 445, 1099

\bibitem[{{Ragan} {et~al.}(2012){Ragan}, {Henning}, {Krause}, {Pitann},
  {Beuther}, {Linz}, {Tackenberg}, {Balog}, {Hennemann}, {Launhardt}, {Lippok},
  {Nielbock}, {Schmiedeke}, {Schuller}, {Steinacker}, {Stutz}, \&
  {Vasyunina}}]{raganEtal2012}
{Ragan}, S., {Henning}, T., {Krause}, O., {et~al.} 2012, \aap, 547, A49

\bibitem[{{Rieke} {et~al.}(2004){Rieke}, {Young}, {Engelbracht}, {Kelly},
  {Low}, {Haller}, {Beeman}, {Gordon}, {Stansberry}, {Misselt}, {Cadien},
  {Morrison}, {Rivlis}, {Latter}, {Noriega-Crespo}, {Padgett}, {Stapelfeldt},
  {Hines}, {Egami}, {Muzerolle}, {Alonso-Herrero}, {Blaylock}, {Dole}, {Hinz},
  {Le Floc'h}, {Papovich}, {P{\'e}rez-Gonz{\'a}lez}, {Smith}, {Su}, {Bennett},
  {Frayer}, {Henderson}, {Lu}, {Masci}, {Pesenson}, {Rebull}, {Rho}, {Keene},
  {Stolovy}, {Wachter}, {Wheaton}, {Werner}, \& {Richards}}]{rieke2004}
{Rieke}, G.~H., {Young}, E.~T., {Engelbracht}, C.~W., {et~al.} 2004, \apjs,
  154, 25

\bibitem[{{Robitaille} {et~al.}(2007){Robitaille}, {Whitney}, {Indebetouw}, \&
  {Wood}}]{robitaille2007}
{Robitaille}, T.~P., {Whitney}, B.~A., {Indebetouw}, R., \& {Wood}, K. 2007,
  \apjs, 169, 328

\bibitem[{{Salatino} {et~al.}(2012){Salatino}, {de Bernardis}, {Masi}, \&
  {Polenta}}]{salatino2012}
{Salatino}, M., {de Bernardis}, P., {Masi}, S., \& {Polenta}, G. 2012, \apj,
  748, 1

\bibitem[{{Skrutskie} {et~al.}(2006){Skrutskie}, {Cutri}, {Stiening},
  {Weinberg}, {Schneider}, {Carpenter}, {Beichman}, {Capps}, {Chester},
  {Elias}, {Huchra}, {Liebert}, {Lonsdale}, {Monet}, {Price}, {Seitzer},
  {Jarrett}, {Kirkpatrick}, {Gizis}, {Howard}, {Evans}, {Fowler}, {Fullmer},
  {Hurt}, {Light}, {Kopan}, {Marsh}, {McCallon}, {Tam}, {Van Dyk}, \&
  {Wheelock}}]{skrutskie2006}
{Skrutskie}, M.~F., {Cutri}, R.~M., {Stiening}, R., {et~al.} 2006, \aj, 131,
  1163

\bibitem[{{Smith}(2006{\natexlab{a}})}]{smith2006}
{Smith}, N. 2006{\natexlab{a}}, \mnras, 367, 763

\bibitem[{{Smith}(2006{\natexlab{b}})}]{smith2006_erratum}
{Smith}, N. 2006{\natexlab{b}}, \mnras, 368, 1983

\bibitem[{{Smith}(2006{\natexlab{c}})}]{smith2006_homunculus}
{Smith}, N. 2006{\natexlab{c}}, \apj, 644, 1151

\bibitem[{{Smith} {et~al.}(2010{\natexlab{a}}){Smith}, {Bally}, \&
  {Walborn}}]{smith2010_jets}
{Smith}, N., {Bally}, J., \& {Walborn}, N.~R. 2010{\natexlab{a}}, \mnras, 405,
  1153

\bibitem[{{Smith} \& {Brooks}(2007)}]{smith2007_census}
{Smith}, N. \& {Brooks}, K.~J. 2007, \mnras, 379, 1279

\bibitem[{{Smith} \& {Brooks}(2008)}]{smith2008}
{Smith}, N. \& {Brooks}, K.~J. 2008, {Handbook of Star Forming Regions, Volume
  II}, ed. {Reipurth, B.}, 138

\bibitem[{{Smith} {et~al.}(2010{\natexlab{b}}){Smith}, {Povich}, {Whitney},
  {Churchwell}, {Babler}, {Meade}, {Bally}, {Gehrz}, {Robitaille}, \&
  {Stassun}}]{smith2010_spitzer}
{Smith}, N., {Povich}, M.~S., {Whitney}, B.~A., {et~al.} 2010{\natexlab{b}},
  \mnras, 406, 952

\bibitem[{{Townsley} {et~al.}(2011){Townsley}, {Broos}, {Corcoran},
  {Feigelson}, {Gagn{\'e}}, {Montmerle}, {Oey}, {Smith}, {Garmire}, {Getman},
  {Povich}, {Remage Evans}, {Naz{\'e}}, {Parkin}, {Preibisch}, {Wang}, {Wolk},
  {Chu}, {Cohen}, {Gruendl}, {Hamaguchi}, {King}, {Mac Low}, {McCaughrean},
  {Moffat}, {Oskinova}, {Pittard}, {Stassun}, {ud-Doula}, {Walborn}, {Waldron},
  {Churchwell}, {Nichols}, {Owocki}, \& {Schulz}}]{townsley2011}
{Townsley}, L.~K., {Broos}, P.~S., {Corcoran}, M.~F., {et~al.} 2011, \apjs,
  194, 1

\bibitem[{{Wang} {et~al.}(2011){Wang}, {Feigelson}, {Townsley}, {Broos},
  {Getman}, {Wolk}, {Preibisch}, {Stassun}, {Moffat}, {Garmire}, {King},
  {McCaughrean}, \& {Zinnecker}}]{wang2011}
{Wang}, J., {Feigelson}, E.~D., {Townsley}, L.~K., {et~al.} 2011, \apjs, 194,
  11

\bibitem[{{Whitney} {et~al.}(2003){Whitney}, {Wood}, {Bjorkman}, \&
  {Wolff}}]{whitney2003}
{Whitney}, B.~A., {Wood}, K., {Bjorkman}, J.~E., \& {Wolff}, M.~J. 2003, \apj,
  591, 1049

\bibitem[{{Whitworth} {et~al.}(1994){Whitworth}, {Bhattal}, {Chapman},
  {Disney}, \& {Turner}}]{whitworth1994}
{Whitworth}, A.~P., {Bhattal}, A.~S., {Chapman}, S.~J., {Disney}, M.~J., \&
  {Turner}, J.~A. 1994, \aap, 290, 421

\bibitem[{{Winston} {et~al.}(2007){Winston}, {Megeath}, {Wolk}, {Muzerolle},
  {Gutermuth}, {Hora}, {Allen}, {Spitzbart}, {Myers}, \& {Fazio}}]{winston2007}
{Winston}, E., {Megeath}, S.~T., {Wolk}, S.~J., {et~al.} 2007, \apj, 669, 493

\bibitem[{{Wolk} {et~al.}(2011){Wolk}, {Broos}, {Getman}, {Feigelson},
  {Preibisch}, {Townsley}, {Wang}, {Stassun}, {King}, {McCaughrean}, {Moffat},
  \& {Zinnecker}}]{wolk2011}
{Wolk}, S.~J., {Broos}, P.~S., {Getman}, K.~V., {et~al.} 2011, \apjs, 194, 12

\bibitem[{{Wright} {et~al.}(2010){Wright}, {Eisenhardt}, {Mainzer}, {Ressler},
  {Cutri}, {Jarrett}, {Kirkpatrick}, {Padgett}, {McMillan}, {Skrutskie},
  {Stanford}, {Cohen}, {Walker}, {Mather}, {Leisawitz}, {Gautier}, {McLean},
  {Benford}, {Lonsdale}, {Blain}, {Mendez}, {Irace}, {Duval}, {Liu}, {Royer},
  {Heinrichsen}, {Howard}, {Shannon}, {Kendall}, {Walsh}, {Larsen}, {Cardon},
  {Schick}, {Schwalm}, {Abid}, {Fabinsky}, {Naes}, \& {Tsai}}]{wright2010}
{Wright}, E.~L., {Eisenhardt}, P.~R.~M., {Mainzer}, A.~K., {et~al.} 2010, \aj,
  140, 1868

\bibitem[{{Yonekura} {et~al.}(2005){Yonekura}, {Asayama}, {Kimura}, {Ogawa},
  {Kanai}, {Yamaguchi}, {Barnes}, \& {Fukui}}]{yonekura2005}
{Yonekura}, Y., {Asayama}, S., {Kimura}, K., {et~al.} 2005, \apj, 634, 476

\end{thebibliography}

\appendix

\section{Online material}

\begin{landscape}
\begin{table}
\caption{Source fluxes as obtained in the \textit{Spitzer} (IRAC: 3.6\,\ensuremath{\upmu\mathrm{m}}, 4.5\,\ensuremath{\upmu\mathrm{m}}, 5.8\,\ensuremath{\upmu\mathrm{m}}, and 8.0\,\ensuremath{\upmu\mathrm{m}}; MIPS: 24\,\ensuremath{\upmu\mathrm{m}}) and \textit{Herschel} (PACS: 70\,\ensuremath{\upmu\mathrm{m}} and 160\,\ensuremath{\upmu\mathrm{m}}, SPIRE: 250\,\ensuremath{\upmu\mathrm{m}}, 350\,\ensuremath{\upmu\mathrm{m}}, and 500\,\ensuremath{\upmu\mathrm{m}}) bands, complemented with WISE photometry from the All-Sky Data Release (3.4\,\ensuremath{\upmu\mathrm{m}}, 4.6\,\ensuremath{\upmu\mathrm{m}}, 12\,\ensuremath{\upmu\mathrm{m}}, and 22\,\ensuremath{\upmu\mathrm{m}}) and JHK\textsubscript{s} photometry as obtained from the 2MASS All-Sky Catalog of Point Sources.}
\label{tab:fluxes}
\centering
\begin{tabular}{lccccccc}
\hline\hline
\noalign{\smallskip}
\multirow{2}{*}{Source} & F\textsubscript{J} & F\textsubscript{H} & F\textsubscript{K\textsubscript{s}} & F\textsubscript{3.4\,\ensuremath{\upmu\mathrm{m}}} & F\textsubscript{3.6\,\ensuremath{\upmu\mathrm{m}}} & F\textsubscript{4.5\,\ensuremath{\upmu\mathrm{m}}} & F\textsubscript{4.6\,\ensuremath{\upmu\mathrm{m}}} \\
                        & [mJy]              & [mJy]              & [mJy]                               & [mJy]                    & [mJy]                    &
[mJy]                    & [mJy]                    \\
\noalign{\smallskip}
\hline
\noalign{\smallskip}
J103427.3--584611 & --                & --                & --                 & 0.939 $\pm 0.015$  & 0.902 $\pm 0.033$  & 2.395 $\pm 0.033$      
                  & 5.173 $\pm 0.023$ \\ 
J103557.6--590046 & 8.624 $\pm 0.069$ & 13.35 $\pm 0.11$  & 12.380 $\pm 0.082$ & 16.095 $\pm 0.061$ & 8.902 $\pm 0.037$  & --
                  & 15.002 $\pm 0.051$ \\ 
J103643.2--583158 & --                & --                & 0.796 $\pm 0.025$  & --                 & 1.585 $\pm 0.026$  & 1.226 $\pm 0.030$
                  & --  \\ 
J103645.9--584258 & 2.245 $\pm 0.019$ & 2.112 $\pm 0.018$ & 1.352 $\pm 0.026$  & --                 & --                 & 0.336 $\pm 0.034$
                  & --   \\ 
J103652.4--583129 & --                & --                & 0.843 $\pm 0.030$  & 4.629 $\pm 0.046$  & 14.790 $\pm 0.030$ & 67.070 $\pm 0.030$
                  & 64.47 $\pm 0.19$   \\
J103700.9--583237 & --                & --                & --                 & 8.596 $\pm 0.067$  & 5.908 $\pm 0.030$  & 17.420 $\pm 0.026$
                  & 26.02 $\pm 0.10$ \\ 
J103703.6--584751 & 0.711 $\pm 0.020$ & 2.610 $\pm 0.036$ & 5.281 $\pm 0.036$  & 5.753 $\pm 0.077$  & 20.540 $\pm 0.029$ & 28.310 $\pm 0.031$
                  & 10.379 $\pm 0.070$ \\
J103726.7--584809 & --                & --                & --                 & --                 & --                 & 0.359 $\pm 0.024$
                  & --  \\
J103737.3--584700 & --                & 0.580 $\pm 0.021$ & 0.873 $\pm 0.022$  & --                 & 1.925 $\pm 0.026$  & 6.835 $\pm 0.029$
                  & --    \\ 
J103739.6--582756 & --                & --                & --                 & --                 & --                 & 3.464 $\pm 0.026$
                  & 4.411 $\pm 0.026$  \\ 
J103741.7--582629 & --                & --                & --                 & --                 & --                 & 0.659 $\pm 0.030$
                  & --     \\ 
J103754.0--584614 & --                & --                & 0.876 $\pm 0.031$  & 5.03 $\pm 0.15$    & 9.175 $\pm 0.029$  & 44.960 $\pm 0.030$
                  & 67.63 $\pm 0.24$ \\
J103804.9--585533 & --                & --                & --                 & 30.11 $\pm 0.17$   & 50.610 $\pm 0.024$ & 158.300 $\pm 0.033$
                  & 211.13 $\pm 0.52$ \\
J103806.6--584002 & --                & 1.119 $\pm 0.033$ & 4.066 $\pm 0.039$  & 8.316 $\pm 0.047$  & 9.277 $\pm 0.024$  & 13.320 $\pm 0.035$
                  & 12.363 $\pm 0.059$ \\ 
J103807.2--584512 & --                & --                & 8.309 $\pm 0.065$  & 92.87 $\pm 0.31$   & 82.700 $\pm 0.025$ & 129.100 $\pm 0.032$
                  & 186.10 $\pm 0.46$  \\ 
J103810.2--584527 & --                & --                & --                 & 7.343 $\pm 0.073$  & 12.330 $\pm 0.024$ & 50.940 $\pm 0.029$
                  & 50.23 $\pm 0.16$ \\
J103842.1--584437 & --                & 2.376 $\pm 0.026$ & 12.26 $\pm 0.074$  & 51.46 $\pm 0.19$   & --                 & 68.850 $\pm 0.030$
                  & 112.14 $\pm 0.29$ \\ 
\hline
\end{tabular}
\tablefoot{The given uncertainties are the individual photometric measurement uncertainties only, as derived (for IRAC and MIPS) or as obtained from the catalogue (for 2MASS and WISE). For \textit{Herschel} we did not obtain an uncertainty from photometry, but use an estimated total uncertainty of 20\%.}
\end{table}
\end{landscape}

\begin{landscape}
\begin{table}
\centering
\ContinuedFloat
\caption{Continued.}
\begin{tabular}{lcccccccccc}
\hline\hline
\noalign{\smallskip}
\multirow{2}{*}{Source} & F\textsubscript{5.8\,\ensuremath{\upmu\mathrm{m}}} & F\textsubscript{8.0\,\ensuremath{\upmu\mathrm{m}}}& F\textsubscript{12\,\ensuremath{\upmu\mathrm{m}}} & F\textsubscript{22\,\ensuremath{\upmu\mathrm{m}}} & F\textsubscript{24\,\ensuremath{\upmu\mathrm{m}}} & F\textsubscript{70\,\ensuremath{\upmu\mathrm{m}}} & F\textsubscript{160\,\ensuremath{\upmu\mathrm{m}}} & F\textsubscript{250\,\ensuremath{\upmu\mathrm{m}}} & F\textsubscript{350\,\ensuremath{\upmu\mathrm{m}}} & F\textsubscript{500\,\ensuremath{\upmu\mathrm{m}}}  \\
                        & [mJy]   & [mJy]   & [mJy]  & [mJy]  & [mJy]   & [Jy]   & [Jy]    & [Jy]    & [Jy]    & [Jy]    \\
\noalign{\smallskip}
\hline
\noalign{\smallskip}
J103427.3--584611 & 2.84 $\pm 0.13$   & 2.21 $\pm 0.11$     & --                & 188.04 $\pm 0.42$ & --                 & 15.8 & 11.0 & 10.7 & 12.2 & 4.47 \\ 
J103557.6--590046 & --                & 70.36 $\pm 0.13$    & 177.29 $\pm 0.27$ & 344.38 $\pm 0.46$ & --                 & 4.03 & 7.63 & 6.09 & --   & --   \\ 
J103643.2--583158 & --                & --                  & --                & --                & --                 & --   & 22.6 & 14.4 & 21.2 & --   \\ 
J103645.9--584258 & --                & --                  & --                & --                & --                 & --   & --   & 9.50 & 8.54 & 2.69 \\ 
J103652.4--583129 & 162.60 $\pm 0.12$ & 197.00 $\pm 0.11$   & 77.32 $\pm 0.39$  & 623.2 $\pm 1.9$   & 755.35 $\pm 0.14$  & 21.3 & 13.9 & 11.2 & --   & 11.2 \\
J103700.9--583237 & --                & --                  & 486.49 $\pm 0.48$ & 1171.2 $\pm 1.3$\tablefootmark{a}  & 395.42 $\pm 0.14$  & --   & 19.1 & 19.9 & 11.2 & 6.96  \\ 
J103703.6--584751 & 35.55 $\pm 0.15$  & 25.51 $\pm 0.13$    & --                & 257.42 $\pm 0.61$ & 181.70 $\pm 0.13$  & 1.32 & 6.72 & 10.8 & 6.32 & 2.89     \\
J103726.7--584809 & --                & --                  & --                & --                & --                 & --   & 12.2 & 11.6 & 9.68 & 7.10  \\
J103737.3--584700 & 17.14 $\pm 0.14$  & 27.08 $\pm 0.11$    & --                & 631.3 $\pm 2.6$   & 149.17 $\pm 0.14$  & --   & 31.4 & 28.1 & 29.8 & 18.4 \\ 
J103739.6--582756 & 7.78 $\pm 0.13$   & 5.47 $\pm 0.13$     & --                & 164.83 $\pm 0.38$ & 245.81 $\pm 0.14$  & --   & 7.68 & 14.1 & 12.0 & 5.68 \\ 
J103741.7--582629 & --                & --                  & --                & --                & --                 & --   & 11.1 & 11.4 & 9.05 & 6.68 \\ 
J103754.0--584614 & 113.80 $\pm 0.13$ & 135.80 $\pm 0.13$   & 134.2 $\pm 1.4$   & 1146.7 $\pm 2.4$  & 858.43 $\pm 0.14$  & 9.03 & 35.8 & 79.6 & 40.7 & 32.4     \\
J103804.9--585533 & 309.70 $\pm 0.14$ & 364.000 $\pm 0.099$ & 81.48 $\pm 0.80$  & 2197.1 $\pm 1.2$  & --                 & 5.91 & 19.8 & 11.8 & 9.01 & 6.23     \\
J103806.6--584002 & 17.23 $\pm 0.12$  & 18.75    $\pm 0.14$ & 60.29 $\pm 0.34$  & 158.6 $\pm 2.2$   & --                 & --   & 10.6 & 13.1 & 5.70 & 2.09 \\ 
J103807.2--584512 & 286.30 $\pm 0.15$ & 526.00   $\pm 0.14$ & 664.2 $\pm 1.1$   & 2409.2 $\pm 1.9$  & 1436.24 $\pm 0.14$ & 15.2 & 35.1 & 26.8 & --   & --   \\ 
J103810.2--584527 & 151.40 $\pm 0.13$ & 245.00   $\pm 0.15$ & 276.9 $\pm 1.1$   & 843.8 $\pm 1.9$   & 1814.13 $\pm 0.14$ & 6.32 & 23.6 & 43.0 & 46.2 & 54.2     \\
J103842.1--584437& 111.60  $\pm 0.11$ & 146.40   $\pm 0.14$ & 269.07 $\pm 0.88$ & 714.2 $\pm 1.4$   & 505.00 $\pm 0.13$  & 24.1 & 14.8 & 15.4 & 10.2 & 5.63 \\ 
\hline
\end{tabular}
\tablefoot{The given uncertainties are the individual photometric measurement uncertainties only, as derived (for IRAC and MIPS) or as obtained from the catalogue (for 2MASS and WISE). For \textit{Herschel} we did not obtain an uncertainty from photometry, but use an estimated total uncertainty of 20\%. \tablefoottext{a}{The extremely high 22\,\ensuremath{\upmu\mathrm{m}} flux compared to the 24\,\ensuremath{\upmu\mathrm{m}} flux is probably due to the sources close proximity to a much brighter point source the contribution of which might not fully have been removed in the WISE All-Sky Data Release.}}
\end{table}
\end{landscape}

\begin{landscape}
\begin{table}
\caption{Model parameters for the sources in Gum\,31 as obtained from the \citet{robitaille2007} models.}
\label{tab:modell}
\centering
\begin{tabular}{c r r r r r r r r c c}
\hline\hline
\noalign{\smallskip}
Source  & \multicolumn{2}{c}{Stellar mass} & \multicolumn{2}{c}{Disk mass} & \multicolumn{2}{c}{Envelope mass} & \multicolumn{2}{c}{Total luminosity} & Best-fit & \multirow{2}{*}{$\displaystyle \frac{\chi^2}{N}$} \\
  & \multicolumn{2}{c}{$[M_\odot]$}  & \multicolumn{2}{c}{$[M_\odot]$} & \multicolumn{2}{c}{$[M_\odot]$} & \multicolumn{2}{c}{$[L_\odot]$} & model &   \\
\noalign{\smallskip}
\hline
\noalign{\smallskip}
J103427.3--584611 & 5.8 & [5.8 -- 7.3] &   1.02\ensuremath{\cdot 10^{-1}} & [4.30\ensuremath{\cdot 10^{-3}} --   1.02\ensuremath{\cdot 10^{-1}}] & 110 & [110 -- 170] & 320 & [320 -- 400] & 3003596 & 6.9
 \\ 
J103557.6--590046 & 1.7 & [1.2 -- 2.0] &   3.91\ensuremath{\cdot 10^{-2}} & [1.76\ensuremath{\cdot 10^{-3}} --   3.91\ensuremath{\cdot 10^{-2}}] & 27  & [4.2 -- 60]  & 42  & [27 -- 68]   & 3015149 & 5.5
 \\
J103645.9--584258 & 4.2 & [3.9 -- 5.2] &   9.61\ensuremath{\cdot 10^{-2}} & [1.09\ensuremath{\cdot 10^{-3}} --   9.61\ensuremath{\cdot 10^{-2}}] & 46  & [18 -- 46]   & 160 & [140 -- 190] & 3006782 & 8.2
 \\ 
J103726.7--584809 & 6.0 & [2.6 -- 6.0] &   2.64\ensuremath{\cdot 10^{-2}} & [1.05\ensuremath{\cdot 10^{-3}} --   1.53\ensuremath{\cdot 10^{-1}}] & 310 & [160 -- 310] & 190 & [94 -- 190]  & 3008699 & 2.6
 \\
J103739.6--582756 & 5.8 & [3.9 -- 6.9] &   2.92\ensuremath{\cdot 10^{-2}} & [2.21\ensuremath{\cdot 10^{-3}} --   2.16\ensuremath{\cdot 10^{-1}}] & 190 & [140 -- 410] & 240 & [160 -- 360] & 3009009 & 4.8
 \\ 
J103741.7--582629 & 3.4 & [2.6 -- 6.0] &   9.53\ensuremath{\cdot 10^{-2}} & [1.62\ensuremath{\cdot 10^{-3}} --   1.53\ensuremath{\cdot 10^{-1}}] & 250 & [160 -- 310] & 94  & [94 -- 190]  & 3005296 & 2.8
 \\ 
J103806.6--584002 & 1.7 & [1.7 -- 2.6] &   4.57\ensuremath{\cdot 10^{-2}} & [1.34\ensuremath{\cdot 10^{-2}} --   4.57\ensuremath{\cdot 10^{-2}}] & 120 & [52 -- 120]  & 38  & [38 -- 58]   & 3016199 & 8.8
 \\  
J103807.2--584512 & 6.6 & [1.7 -- 7.3] &   6.77\ensuremath{\cdot 10^{-4}} & [6.77\ensuremath{\cdot 10^{-4}} --   5.06\ensuremath{\cdot 10^{-1}}] & 130 & [5.3 -- 310] & 890 & [220 -- 890] & 3010777 & 7.8
 \\ 
J103842.1--584437 & 3.5 & [1.7 -- 3.9] &   1.03\ensuremath{\cdot 10^{-2}} & [6.45\ensuremath{\cdot 10^{-3}} --   1.97\ensuremath{\cdot 10^{-1}}] & 220 & [13 -- 220]  & 170 & [130 -- 190] & 3011717 & 6.6
 \\ 

\noalign{\smallskip}
\hline
\end{tabular}
\tablefoot{For every model parameter the best-fit-value is given in the respective first column, followed by a range defined by the minimum and maximum value obtained from models constrained by a $\chi^2$ criterion. The second-to-last and last columns give the identifier of the best-fit model and its $\chi^2\!/N$ (with $N$ representing the number of data points).}
\end{table}
\end{landscape}

\end{document}